\documentclass[nofootinbib,a4paper,aps,prd,10pt,superscriptaddress,showkeys,twocolumn]{revtex4}
\usepackage{graphicx}
\usepackage{amsfonts}
\usepackage{amssymb} 
\usepackage{amsmath}
\usepackage{float}
\usepackage{natbib}
\usepackage{dcolumn}
\usepackage{bm}
\usepackage{latexsym,color}
\usepackage{setspace}

\usepackage{hyperref}
\hypersetup{%
    ,urlcolor=blue
    ,citecolor=blue
    ,linkcolor=blue
    }

\begin{document}

\title{Shirokov and Shapiro Effects in the Hartle-Thorne Spacetime}

\author{Anuar~\surname{Idrissov}}
\email[]{anuar.idrissov@gmail.com}
\affiliation{Instituto de Ciencias Nucleares, Universidad Nacional Aut\`onoma de M\`exico, Mexico}
\affiliation{Fesenkov Astrophysical Institute, Observatory 23, 050020 Almaty, Kazakhstan}
\affiliation{Al-Farabi Kazakh National University, Al-Farabi Ave., 71, Almaty, 050040, Kazakhstan}

\author{Kuantay~\surname{Boshkayev}}
\email[]{kuantay@mail.ru}
\affiliation{Al-Farabi Kazakh National University, Al-Farabi Ave., 71, Almaty, 050040, Kazakhstan}

\author{Serzhan~Momynov}
\email[]{s.momynov@gmail.com}
\affiliation{Al-Farabi Kazakh National University, Al-Farabi Ave., 71, Almaty, 050040, Kazakhstan}
\affiliation{Satbayev University, Satbayev St., 22, Almaty, 050013, Kazakhstan}

\author{Hernando~\surname{Quevedo}}
\email[]{quevedo@nucleares.unam.mx}
\affiliation{Instituto de Ciencias Nucleares, Universidad Nacional Aut\`onoma de M\`exico, Mexico}
\affiliation{Al-Farabi Kazakh National University, Al-Farabi Ave., 71, Almaty, 050040, Kazakhstan}
\affiliation{Dipartimento di Fisica and ICRA, Universit\`a di Roma “La Sapienza”, Roma, Italy}

\author{Daniya~\surname{Utepova}}
\email[]{utepova.daniya@gmail.com}
\affiliation{Abai Kazakh National Pedagogical University, Dostyk Ave., 13, Almaty, 050010, Kazakhstan}
\affiliation{Al-Farabi Kazakh National University, Al-Farabi Ave., 71, Almaty, 050040, Kazakhstan}

\author{Ainur~\surname{Urazalina}}
\email[]{y.a.a.707@mail.ru}
\affiliation{Al-Farabi Kazakh National University, Al-Farabi Ave., 71, Almaty, 050040, Kazakhstan}

\author{Bagila~Baitimbetova}
\email[]{baitim@physics.kz}
\affiliation{Satbayev University, Satbayev St., 22, Almaty, 050013, Kazakhstan}

\date{\today}

\begin{abstract}
We investigate the influence of rotation and quadrupole deformations of astrophysical compact objects on the Shirokov and Shapiro effects within the Hartle-Thorne spacetime, which describes the exterior gravitational field of slowly rotating, slightly deformed celestial objects. Using geodesic deviation equations, we analyze the oscillatory motion of neighboring test particle trajectories and show how the combined impact of angular momentum $J$ and quadrupole moment $Q$ affects the Shirokov effect. The results are compared with our previous analysis for the Lense-Thirring and Zipoy-Voorhees metrics, revealing consistent trends in the coupling between radial and azimuthal oscillations. 
Importantly, by evaluating the period splitting in the weak-field regime we
show that the dominant contribution to the Shirokov effect is the Newtonian
quadrupole moment of the source rather than the relativistic mass term
originally identified by Shirokov. 
For the Shapiro time delay, we examine two limiting configurations: (i) the Lense-Thirring frame -- dragging case with $J^2=0$, $Q=0$ and $J\neq0$, where the effect persists for both positive and negative values of the angular momentum; and (ii) the static quadrupolar case with  $J=0$ and $Q\neq0$, where more oblate sources produce a stronger gravitational time delay with increasing distance. We also study these effects in the Hartle-Thorne spacetime without employing the weak-field approximation, performing a full numerical analysis. In particular, we examine the mimicking effects produced by the quadrupole deformation and the angular momentum of the compact object. These results illustrate how the deformation and rotation of compact objects influence the relativistic observables in the surrounding spacetime. 
\end{abstract}


\maketitle 
 
\section{Introduction} \label{sec:intro}

Neutron stars and hypothetical quark stars are among the densest objects in the
Universe: they pack one to two solar masses into a sphere of roughly
$12$--$15\,$km, at densities above nuclear, where the gravitational field is
strong enough that general relativistic effects govern the motion of test
particles \cite{ShapiroTeukolsky1983, Haensel2007, 2013ApJ...765L...5S, 2016ARA&A..54..401O}.
This makes them natural laboratories for strong-field gravity and for the
behaviour of dense matter \cite{Haensel2007, Lattimer2012}. Unlike black holes,
they have a finite surface, and rotation, tidal forces, and strong magnetic
fields make their shape deviate from spherical symmetry
\cite{2013PhRvD..88b3009Y,1999A&A...352..211K,2000A&A...356..234K,2015JCAP...05..016D,2025PhRvD.112b3030K}.
These deviations are encoded in the star's multipole moments, which depend on
its mass $M$, angular momentum $J$, and quadrupole moment $Q$
\cite{Lattimer2016,Lo2013,Pappas2012,PappasApos2012}. Modelling them therefore
requires a metric that includes both rotation and quadrupolar deformation. In this work, 
we focus on deformation caused by rotation.

The Hartle-Thorne (HT) metric provides such a description. It is an approximate,
stationary, axially symmetric solution of Einstein's equations, obtained as a
second-order expansion in the star's angular velocity
\cite{Hartle1967,1968ApJ...153..807H}, and characterised by three independent
parameters $M$, $J$, and $Q$. In the Kerr metric the multipole moments obey the
no-hair relation $M_\ell + i S_\ell = M(ia)^\ell$, which fixes the quadrupole to
$Q_{\rm Kerr}=J^{2}/M$; in the exterior HT metric, by contrast, $Q$ and $J$ are
independent. This freedom is particularly useful for modeling neutron stars, white
dwarfs, and other slowly rotating compact objects
\cite{2014NuPhA.921...33B, Frutos2015,Boshkayev2016,Pappas2017}. Many exact
solutions also describe rotating, deformed objects -- for example the
Quevedo-Mashhoon, rotating $q$-metric, Manko, and Pachon-Rueda metrics
\cite{1989PhRvD..39.2904Q,1991PhRvD..43.3902Q,2000PhRvD..61h1501M,2006PhRvD..73j4038P,2012PhRvD..86f4043B,2018RSOS....580640F,2020CQGra..37e5006A}
-- but most observed neutron stars rotate slowly and deform only weakly, so the
HT metric provides an adequate approximation.

Because of this flexibility, the HT metric has been applied to circular geodesic
motion \cite{2013NCimC..36S..31B}, accretion disk dynamics
\cite{2024EPJP..139..273B}, quasi-periodic oscillations
\cite{2025arXiv250611581B}, light deflection in plasma
\cite{2023PhRvD.108h4043B}, and gravitational lensing and other relativistic
effects near compact objects
\cite{Quevedo2011,Quevedo2013,Manko2000,2019JPhCo...3h5018A,2025GReGr..57...55M}.
Such studies connect theoretical models to observations and help constrain the
neutron star equation of state.

The geodesic deviation equation provides a useful tool for studying the relative motion of nearby geodesics in curved spacetime
\cite{misner1973gravitation}. It is well suited to the \textit{Shirokov effect}:
the difference between the radial and vertical oscillation frequencies of a test
particle near a circular orbit, produced by spacetime curvature
\cite{shirokov1973one}. Shirokov first studied this in the Schwarzschild field;
it was later extended to the Lense-Thirring metric \cite{nduka1977shirokov},
where rotation adds a phase shift, to the Kerr metric
\cite{vladimirov1981small}, and to post-Newtonian and static axisymmetric fields
\cite{melkumova1990calculation}. More recently, work in the Zipoy-Voorhees
($q$-)metric \cite{voorhees1970static,zipoy1966topology} showed that quadrupole
deformation strongly affects the oscillatory motion and the frequency difference
between neighbouring trajectories \cite{Utepova2025}.

A second key observable is the \textit{Shapiro time delay}
\cite{Shapiro1964,Klioner2015}, the extra travel time of light passing near a
massive body. First measured by radar ranging to Venus past the Sun
\cite{d1992introducing} and later tested by the Cassini mission
\cite{bertotti2003test}, it has since been extended to rotating geometries
\cite{dymnikova1986gravitational} and to higher-order and moving source
configurations, and is now a standard probe in pulsar timing and gravitational
lensing. In our earlier $q$-metric analysis \cite{Utepova2025} we found that
quadrupole deformation markedly changes the magnitude and distance dependence of
the delay. Quantum corrections to the delay around a spinning source have also
been studied within effective field theory \cite{2017CQGra..34p5008B}.

In this work, we extend these analyzes to the Hartle-Thorne spacetime. Using the
geodesic deviation formalism we derive the Shirokov effect and examine how
rotation and quadrupole deformation together shape the oscillation frequencies
and phase differences, and we compute the Shapiro delay for light propagation
and its dependence on $J$ and $Q$. For the delay we consider a strong-field
configuration of two test bodies on circular orbits around a neutron star
(Fig.~\ref{fig:scheme}). Such multi-body systems are astrophysically motivated:
planetary mass companions are confirmed around pulsars such as PSR~B1257+12
\cite{WolszczanFrail1992}, and there is evidence for fallback disks and debris
around neutron stars \cite{PatrunoKama2017,ShannonCordes2013}, with small rocky
bodies able to survive at close orbits \cite{KoteraMottez2016} down to the
relativistic Roche limit \cite{Gourgoulhon2019}. We treat this delay as a
controlled strong-field analogue of the classical Solar-System experiment rather
than a direct observational prediction, while noting its potential relevance to
precise timing of pulsars and X-ray binaries
\cite{Pappas2017,Quevedo2011,Lattimer2016}.

The novelty of the present work can be summarised as follows:
\begin{itemize}
\item[(i)] We provide the complete geodesic deviation system in the exterior
Hartle-Thorne spacetime, with all coefficients given in analytic form through
second order in the spin and first order in the quadrupole moment.
\item[(ii)] We identify the Shirokov observable as the reparametrization invariant
ratio of the vertical and in-plane oscillation periods, and the secular vertical
displacement it generates, as distinct from the individual epicyclic frequencies,
which are well known.
\item[(iii)] We show that, in the weak-field regime, the dominant contribution to
this splitting is the Newtonian quadrupole moment of the source -- not the
relativistic mass term emphasised by Shirokov -- a point developed further in a
companion paper~\cite{idrissov2026newtonian}.
\item[(iv)] We derive an analytic Shapiro delay in the Hartle-Thorne exterior
through second order in spin and first order in quadrupole, with consistent
Lense-Thirring and $q$-metric limits, and we quantify the spin-quadrupole
degeneracy (mimicking) common to both observables.
\end{itemize}

This paper is organized as follows. Section~\ref{sec:geoddev} introduces the
Hartle-Thorne background and derives the geodesic deviation equations.
Section~\ref{sec:shir} analyses the Shirokov effect -- the oscillation-period
splitting of neighbouring circular geodesics -- and its dependence on $J$ and
$Q$, including the parametrization-invariant observable, the spin-quadrupole
degeneracy, and the weak-field limit. Section~\ref{sec:shap} derives the Shapiro
time delay in the same spacetime, together with its Lense-Thirring,
static-quadrupole, and $q$-metric limits and its relation to existing timing
formulae. Section~\ref{conclusions} summarises the results and their
implications. Throughout we use geometrized units $G=c=1$ and signature
$(-,+,+,+)$.

\section{Geodesic deviation in the Hartle-Thorne spacetime} \label{sec:geoddev}

The line element for the exterior Hartle-Thorne metric is given by \cite{1968ApJ...153..807H} \\

\begin{widetext}
\begin{align}\label{ht1}
ds^2=\,&-\left(1-\frac{2{ M }}{r}\right) \left[1+2k_1P_2(\cos\theta)+2\left(1-\frac{2{M}}{r}\right)^{-1} \frac{J^{2}}{r^{4}}(2\cos^2\theta-1)\right]dt^2+\left(1-\frac{2{M}}{r}\right)^{-1}\nonumber\\
&\times\left[1-2k_2P_2(\cos\theta)-2\left(1-\frac{2{M}}{r}\right)^{-1}\frac{J^{2}}{r^4}\right]dr^2
+r^2[1-2k_3P_2(\cos\theta)](d\theta^2+\sin^2\theta d\phi^2) -\frac{4J}{r}\sin^2\theta dt d\phi\,,
\end{align}
\end{widetext}
where the functions $k_1$, $k_2$, and $k_3$ are defined as
\begin{eqnarray}\label{k123}
k_1&=&\frac{J^{2}}{{M}r^3}\left(1+\frac{{M}}{r}\right)+\frac{5}{8}\frac{Q-J^{2}/{M}}{{M}^3}Q_2^2\left(\frac{r}{{M}}-1\right),\nonumber\\
k_2&=&k_1-\frac{6J^{2}}{r^4},\nonumber\\
k_3&=&k_1+\frac{J^{2}}{r^4}+\frac{5}{4}\frac{Q-J^{2}/{M}}{{M}^2\left(r^2-2Mr\right)^{1/2}} Q_2^1\left(\frac{r}{M}-1\right)\,,\nonumber
\end{eqnarray}
with
\begin{eqnarray}\label{ht2}
P_{2}(\cos\theta)&=&\frac{1}{2}(3\cos^{2}\theta-1),\nonumber\\
Q_{2}^{1}(x)&=&(x^{2}-1)^{1/2}\left[\frac{3x}{2}\ln\frac{x+1}{x-1}-\frac{3x^{2}-2}{x^{2}-1}\right],\nonumber\\
Q_{2}^{2}(x)&=&(x^{2}-1)\left[\frac{3}{2}\ln\frac{x+1}{x-1}-\frac{3x^{3}-5x}{(x^{2}-1)^2}\right],\nonumber
\end{eqnarray}
where $x=r/M-1$ and $M$, $J$, and $Q$ denote the total mass, angular momentum, and mass quadrupole moment of the rotating source, respectively. Here, $P_{2}(x)$ is the Legendre polynomial of the first kind, while $Q_l^m$ are the associated Legendre functions of the second kind. The parameters $J\sim\Omega_{Star}$ and $Q\sim\Omega_{Star}^2$ are related to the angular velocity  $\Omega_{Star}$ of the rotating body that acts as the gravity source. Furthermore, we use the following notation
\begin{align} \nonumber
t=x^{0}, \quad r =x^{1},\quad\theta=x^{2},\quad \phi=x^{3}.
\end{align}
The Shirokov effect can be studied by analyzing the relative motion of neighboring geodesics, governed by the geodesic deviation equation \cite{misner1973gravitation}
\begin{equation}\label{eq:GDE}
    \frac{D^2 \xi^{\alpha}}{d s^{2}} - R^\alpha_{\mu\nu\rho} u^{\mu} u^{\nu} \xi^{\rho} = 0,
\end{equation}
where  $D/d s$ is the covariant derivative, $ R^\alpha_{\mu\nu\rho}$ is the Riemann tensor, $u^{\mu}=d x^{\mu}/d s$ is the 4-velocity tangential to the geodesic and the geodesic deviation vector is denoted as  $\xi^{\alpha}$.

Following the original approach of Shirokov \cite{shirokov1973one}, the deviation equations can be equivalently expressed in terms of the Christoffel symbols as ~\cite{OhanianRuffini2013}
\begin{equation}\label{dev1}
\frac{d^{2}\xi^{i}}{ds^{2}}+2\Gamma^{i}_{jk}u^{j}\frac{d\xi^{k}}{ds}+\frac{\partial{\Gamma^{i}_{jk}}}{\partial{x}^{l}}u^{j}u^{k}\xi^{l}=0.
\end{equation}
By substituting the non-zero connection coefficients of the Hartle–Thorne metric \eqref{ht1}, one obtains the following system of coupled equations for the deviation components $\xi^{i}$:
\begin{equation}\label{eq2}
\frac{d^{2}\xi^{1}}{ds^{2}}+a_1\frac{d\xi^{3}}{ds}+a_2\frac{d\xi^0}{ds}+a_3\xi^1=0,
\end{equation}
\begin{equation}\label{eq3}
\begin{gathered}
\frac{d^{2}\xi^{2}}{ds^{2}}+h \xi^2=0,
\end{gathered}
\end{equation}
\begin{equation} \label{eq4}
\begin{gathered}
\frac{d^{2}\xi^{3}}{ds^{2}}+b\frac{d\xi^1}{ds}=0,
\end{gathered}
\end{equation}
\begin{equation}\label{eq5}
\begin{gathered}
\frac{d^{2}\xi^{0}}{ds^{2}}+c\frac{d\xi^1}{ds}=0.
\end{gathered}
\end{equation}
Here, the coefficients $a_1$, $a_2$, $a_3$, $b$, $c$ and $h$ encode the effects of mass, rotation, and quadrupole deformation on the geodesic deviation. Each of these coefficients can be expressed as a series expansion in terms of the angular momentum $j=J/M^2$ and the quadrupole moment $q=Q/M^3$, and by employing following definition  $f=1 - 2 M/r$:
\begin{equation}\label{eq:abc-expansion}
\begin{aligned}
a_i &= a_{i0}\!\left(1 + j\,a_{i1} + j^{2} a_{i2} + q\,a_{i3}\right),\\[4pt]
b   &= b_{0}\!\left(1 + j\,b_{1} + j^{2} b_{2} + q\,b_{3}\right),\\[4pt]
c   &= c_{0}\!\left(1 + j\,c_{1} + j^{2} c_{2} + q\,c_{3}\right),\\[4pt]
h   &= h_{0}\!\left(1 + j\,h_{1} + j^{2} h_{2} + q\,h_{3}\right).
\end{aligned}
\end{equation}
The explicit forms of the coefficients $a_{1}$, $a_{2}$, $a_{3}$, $b$, $c$, and
$h$, expanded through $\mathcal{O}(j^{2})$ and $\mathcal{O}(q)$, are collected
in Appendix~\ref{app:gde_coeffs} for completeness.

\section{Shirokov effect} \label{sec:shir}
In his pioneering work of relativistic orbital dynamics, Shirokov \cite{shirokov1973one} introduced a simplified physical model to illustrate the motion of a test particle in curved spacetime. He considered a hollow spherical satellite whose center of mass follows a circular geodesic in the field of the Earth, using the Schwarzschild metric. A small test particle, placed at the center of this sphere, is then given a tiny initial displacement and velocity relative to the satellite's frame. The subsequent motion of the particle represents small oscillations about the reference geodesic. When radial or vertical perturbations are introduced, the oscillations occur with slightly different characteristic periods
\begin{equation}\label{difperiods}
\Delta T = T_\theta - T_r \ (\text{or } T_\phi) \approx -\frac{3M}{r} T_0 ,
\end{equation}
where $T_0$ is the corresponding Newtonian period. The appearance of this small frequency difference, absent in Newtonian gravity, directly reflects spacetime curvature and provides a subtle relativistic correction that, in principle, can be measured with modern experimental precision \cite{will2014confrontation}.

The radial and vertical epicyclic frequencies of circular geodesics in the
Hartle-Thorne spacetime are well established and have been used extensively in the modelling of accretion disc oscillations and quasi-periodic oscillations (see, e.g., \cite{2001A&A...374L..19A,2001AcPPB..32.3605K, 2022arXiv220310653M, 2025arXiv250611581B, Urbancova:2019ep}). Likewise, the use of the geodesic deviation equation, including higher-order treatments, is a standard route to such frequencies \cite{Kerner:2001ep,Colistete:2002km}. Our aim here is not to rederive these well-known frequencies, but to construct from them the Shirokov observable -- the relative phase drift between the vertical and in-plane oscillations of neighboring geodesics -- and to determine its dependence on the angular momentum and quadrupole moment.
\subsection{Circular geodesics and oscillation frequencies}\label{sec:shir_freq}
Here, we generalize Shirokov's approach to the strong-field regime near a rotating and deformed compact object described by the Hartle-Thorne metric. Instead of a terrestrial satellite, we consider a test particle in a slightly perturbed circular orbit around a slowly rotating neutron star, whose oscillations about the reference geodesic result from the combined influence of spacetime curvature, frame dragging, and quadrupole deformation. The resulting differences between the azimuthal, radial, and vertical oscillation periods define the relativistic extension of the Shirokov effect in the Hartle-Thorne spacetime~\cite{2013MNRAS.433.1903U}.

To generalize this approach to an axially symmetric spacetime, we now consider the metric (\ref{ht1}) and the geodesic equations  
\begin{align}\label{geodeq}
\frac{du^{i}}{ds}+\Gamma^{i}_{jk}u^{j}u^{k}=0,
\end{align}
with $u^{1}=dr/ds=0$ and $u^{2}=d\theta/ds=0$. Assuming circular motion in the equatorial plane, only the components \( u^0 \) and \( u^3 \) of the 4-vector velocity are non-zero. Substituting these values and  the non-vanishing Christoffel symbols in the geodesic equation for the $ \phi- $component, we obtain 
\begin{align} \label{explgeodeq_compact}
&\frac{f\!\left[M (u^{0})^{2} - r^{3} (u^{3})^{2}\right]}{r^{2}}
+ j\,K(r)\,u^{0}u^{3} \nonumber\\
&\qquad\qquad + j^{2}\!\left[C_{0}(r)\,(u^{0})^{2} + C_{3}(r)\,(u^{3})^{2}\right] \nonumber\\
&\qquad\qquad + q\!\left[D_{0}(r)\,(u^{0})^{2} + D_{3}(r)\,(u^{3})^{2}\right] = 0.
\end{align}
The functions $K$, $C_{0}$, $C_{3}$, $D_{0}$, and $D_{3}$ are listed in
Appendix~\ref{app:circ_coeffs}.
This provides a dynamical relation between $ u^0 $ and $ u^3 $.
Additionally, the normalization condition \( g_{\mu \nu} u^\mu u^\nu = -1 \) yields
\begin{align} \label{normaleq}
-f\,(u^{0})^{2} 
&+ r^{2}(u^{3})^{2} 
- j\,\frac{4 M^{2}}{r}\,u^{0}u^{3} \nonumber\\
&+ j^{2}\!\left[A_{0}(r)\,(u^{0})^{2} + A_{3}(r)\,(u^{3})^{2}\right] \nonumber\\
&+ q\!\left[B_{0}(r)\,(u^{0})^{2} + B_{3}(r)\,(u^{3})^{2}\right] = -1.
\end{align}
The functions $A_{0}$, $A_{3}$, $B_{0}$, and $B_{3}$ are given in
Appendix~\ref{app:circ_coeffs}.
Solving the equations given above, we determine the approximate values of $u^{3}$ and  $u^{0}$ as follows
\begin{equation}\label{equ0}
\begin{gathered}
u^{0}
= u^{01}\left(
1
+ j\,g^{00}
+ j^{2} g^{1}
+ q\,g^{2}
\right),
\end{gathered}
\end{equation}
\begin{equation}\label{equ3}
\begin{gathered}
u^{3}=u^{03}\left(1+ j d^{03} +j^{2} d^{3}+q d^{4}\right),
\end{gathered}
\end{equation}
with $u^{01}=(1-3M/r)^{-1/2}$ and $u^{03}=M^{1/2}/[r(r-3M)^{1/2}]$; the
remaining coefficients $g^{00},g^{1},g^{2}$ and $d^{03},d^{3},d^{4}$ are given
in Appendix~\ref{app:circ_coeffs}.

Substituting \eqref{equ0} and \eqref{equ3} in \eqref{hcoef},  we obtain the explicit value of $h$, which for further use we define as $h:=\Omega^2$. Then, 
\begin{equation}\label{bigfreq}
\begin{gathered}
\Omega^{2}=(\Omega_{01})^{2} \left[1+j \Omega_{02}+j^{2} \Omega_{03}+q \Omega_{04} \right],
\end{gathered}
\end{equation}
where $(\Omega_{01})^{2}=M/[r^{2}(r-3M)]$ and
$\Omega_{02}=-6M^{3/2}f/[r^{1/2}(r-3M)]$; the second-order spin and quadrupole
coefficients $\Omega_{03}$ and $\Omega_{04}$ are given in
Appendix~\ref{app:freq_coeffs}.
Then, Eq. \eqref{eq3} becomes
\begin{equation}\label{eq18}
\begin{gathered}
\frac{d^{2}\xi^{2}}{ds^{2}}+\Omega^{2}\xi^{2}=0,
\end{gathered}
\end{equation}
whose solution is
\begin{equation}\label{eq19}
\begin{gathered}
\xi^{2}=\xi_{0}^{2} e ^{i\Omega s}.
\end{gathered}
\end{equation}

Furthermore, the solutions of Eqs. \eqref{eq2}, \eqref{eq4}, and \eqref{eq5}  can be written in the following form
\begin{equation}\label{eq20}
\begin{gathered}
\xi^{0}=\xi_{0}^{0} e ^{i\omega s}, ~\xi^{1}=\xi_{0}^{1} e ^{i\omega s},~\xi^{3}=\xi_{0}^{3} e ^{i\omega s}.
\end{gathered}
\end{equation}
Substituting
\eqref{eq20} in \eqref{eq2}, \eqref{eq4}, and \eqref{eq5} produces the following homogeneous linear system of equations
\begin{equation}
\begin{cases}
\;(a_3 - \omega^2) \xi_0^1 + i\omega a_1 \xi_0^3 + i\omega a_2 \xi_0^0 = 0, \\[4pt]
\;i\omega b \xi_0^1 - \omega^2 \xi_0^3 = 0, \\[4pt]
\;i\omega c \xi_0^1 - \omega^2 \xi_0^0 = 0.
\end{cases}
\label{eq:system}
\end{equation}
Non-trivial solutions require the determinant of the coefficient matrix to vanish, leading to two distinct frequency solutions, first: $\omega^2 = 0$ and second:
\begin{equation} \label{smallfreq}
\begin{gathered}
\omega^2 =(\omega_{01})^2 \left[1+j \omega_{02}+j^{2} \omega_{03}+q \omega_{04} \right].
\end{gathered}
\end{equation}
where $(\omega_{01})^{2}=M(r-6M)/[r^{3}(r-3M)]$ and
$\omega_{02}=6M^{3/2}r^{1/2}f/[(r-3M)(r-6M)]$; the coefficients $\omega_{03}$
and $\omega_{04}$ are given in Appendix~\ref{app:freq_coeffs}.
\subsection{The Shirokov observable and its parametrization invariance}\label{sec:shir_invariance}

It is important to emphasize that $\Omega$ and $\omega$ defined above are frequencies with
respect to the proper time $s$ along the unperturbed circular geodesic. They are
related to the coordinate-time (distant-observer) frequencies by the common
redshift/time-dilation factor $u^{t}$,
\begin{equation}
\Omega_{(t)} = \frac{\Omega}{u^{t}}, \qquad
\omega_{(t)} = \frac{\omega}{u^{t}},
\end{equation}
which allows direct comparison with the coordinate-time epicyclic frequencies
employed in the accretion disc and QPO literature. We stress, however, that the
quantities reported below -- the period ratio $T_r/T_\theta$, the relative shift
$\Delta T/T_0$, and the accumulated vertical displacement $\xi^{2}(n)$ -- are
\emph{independent of this choice of parametrization}, since the common factor
$u^{t}$ cancels:
\begin{equation}
\frac{T_r^{(t)}}{T_\theta^{(t)}}
= \frac{2\pi u^{t}/\omega}{2\pi u^{t}/\Omega}
= \frac{\Omega}{\omega}
= \frac{T_r}{T_\theta}.
\end{equation}
The Shirokov effect, therefore, is a parametrization-independent
geometric property of neighboring worldlines, rather than a relabeling of the
known coordinate-time frequencies.

To obtain the oscillation periods, we substitute the frequencies $\Omega$ and $\omega$ into the definitions $T_{\theta} = 2\pi / \Omega$ and $T_{r} = T_{\phi} = 2\pi / \omega$. Since our primary interest is the strong-field regime relevant for neutron stars, the periods are evaluated using the full expressions \eqref{bigfreq} and \eqref{smallfreq}, giving
\begin{equation}\label{TroverTtheta}
\begin{gathered}
\frac{T_r}{T_\theta}=\left(\frac{T_r}{T_\theta}\right)_{01} \left[1+j \left(\frac{T_r}{T_\theta}\right)_{02}+j^{2} \left(\frac{T_r}{T_\theta}\right)_{03}+q \left(\frac{T_r}{T_\theta}\right)_{04} \right],
\end{gathered}
\end{equation}
where the leading coefficients are
$(T_r/T_\theta)_{01}=(1-6M/r)^{-1/2}$ and
$(T_r/T_\theta)_{02}=-6(r-2M)(M/r)^{3/2}/(r-6M)$; the second-order coefficient
$(T_r/T_\theta)_{03}$ and $(T_r/T_\theta)_{04}$ are given in
Appendix~\ref{app:freq_coeffs}.
\subsection{Period splitting and vertical displacement}\label{sec:shir_strongfield}
To facilitate comparison with the $q$-metric and Lense-Thirring limits, we expand to $\mathcal{O}(r^{-5})$ and $\mathcal{O}(M^{2})$, keeping all contributions explicit in $J$, $J^{2}$, and $Q$; all numerical evaluations and plots, however, use the full, non-expanded expressions. We obtain
\begin{align}
\Omega &\approx
\sqrt{\frac{M}{r^3}}\Bigg[
1 + \frac{3M}{2r} + \frac{27M^{2}}{8r^{2}} \nonumber\\
&\qquad\qquad\qquad 
- \frac{3J\big(63 M^{2} + 20 Mr + 8r^{2}\big)}{8\sqrt{M}\,r^{7/2}} \nonumber\\
&\qquad\qquad\qquad
- \frac{5 J^{2}}{2M r^{3}}
+ \frac{Q\big(71M + 18r\big)}{8M r^{3}}
\Bigg],
\label{eq:Omega_HT}
\end{align}
\begin{align}
\omega &\approx 
\sqrt{\frac{M}{r^3}}\Bigg[
1 - \frac{3M}{2r} - \frac{45M^{2}}{8r^{2}} \nonumber\\
&\qquad\qquad\qquad
+ \frac{3J\big(231M^{2}+44Mr+8r^{2}\big)}{8\sqrt{M}\,r^{7/2}} \nonumber\\
&\qquad\qquad\qquad
- \frac{11J^{2}}{2M r^{3}}
- \frac{Q\big(103M+6r\big)}{8M r^{3}}
\Bigg],
\label{eq:omega_HT}
\end{align}
\begin{align}
T_{\theta} &\approx
T_{0}\Bigg[
1 - \frac{3M}{2r}
- \frac{9M^{2}}{8r^{2}}
\nonumber\\[3pt]
&\quad
- \frac{3J(M - 2r)}{2\sqrt{M} r^{5/2}}
+ \frac{23J^{2}}{2M r^{3}}
- \frac{Q(17M + 18r)}{8M r^{3}}
\Bigg],\\[8pt]
T_{r} &\approx
T_{0}\Bigg[
1 + \frac{3M}{2r}
+ \frac{63M^{2}}{8r^{2}}
\nonumber\\[3pt]
&\quad
- \frac{3J(17M + 2r)}{2\sqrt{M}\,r^{5/2}}
+ \frac{29J^{2}}{2M r^{3}}
+ \frac{Q(121M + 6r)}{8M r^{3}}
\Bigg],
\label{eq:Ttheta_Tr}
\end{align}
where $T_{0}=2\pi/\sqrt{\frac{M}{r^{3}}}$ is the Newtonian period. The difference between 
$T_{\theta}$ and $T_{r}$ (or $T_{\phi}$) is defined as
\begin{align}
\Delta T &= T_{\theta} - T_{r}
\nonumber\\[3pt]
&\approx
T_{0}\Bigg[
-\frac{3M}{r}
- \frac{9M^{2}}{r^{2}}
+ \frac{6J(4M + r)}{\sqrt{M}\,r^{5/2}} \nonumber\\
&\qquad\qquad\qquad- \frac{3J^{2}}{M r^{3}}
- \frac{3Q(23M + 4r)}{4M r^{3}}
\Bigg].
\label{eq:DeltaT}
\end{align}

\begin{figure*}
{\hfill
\includegraphics[width=8cm]{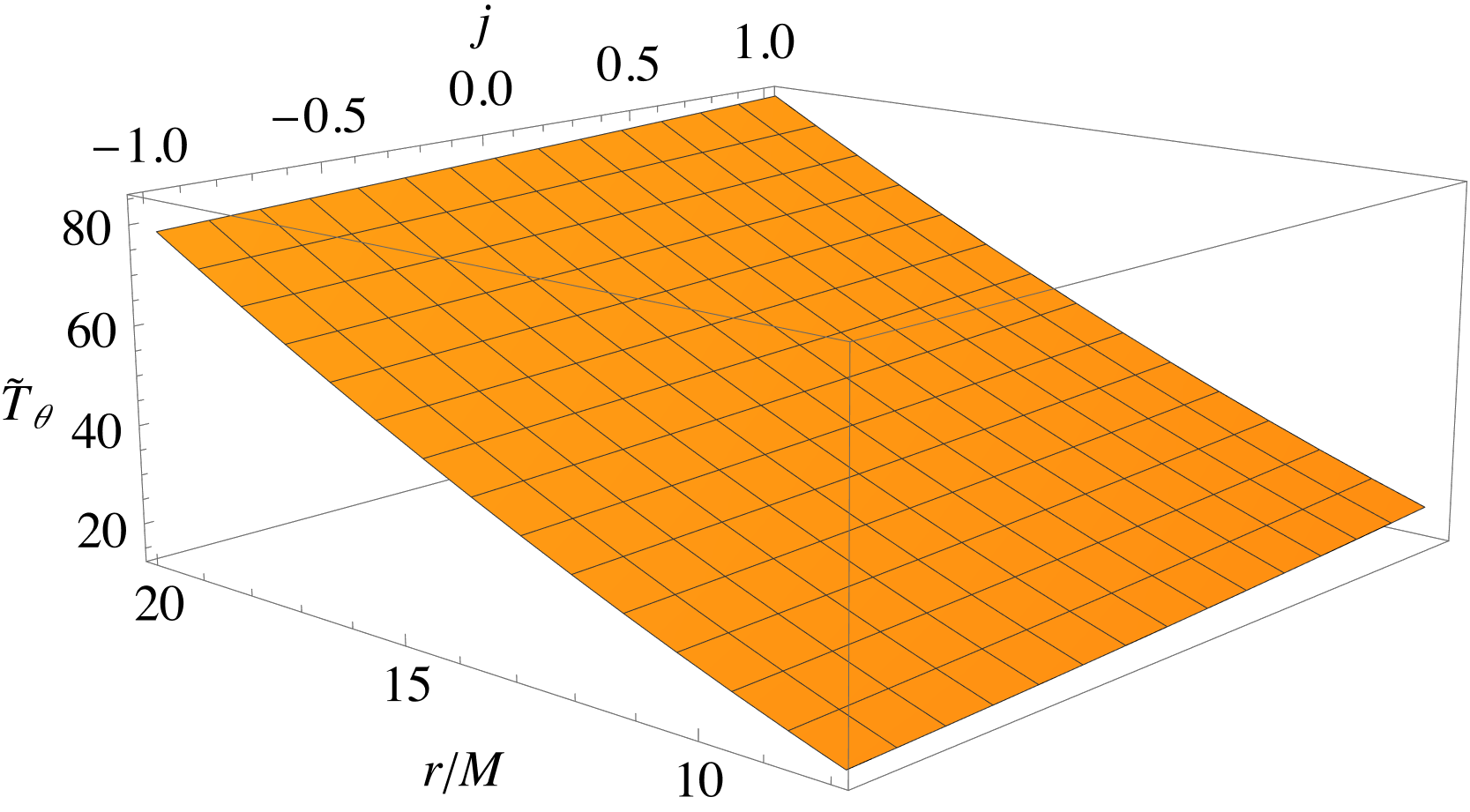}\hfill
\includegraphics[width=8cm]{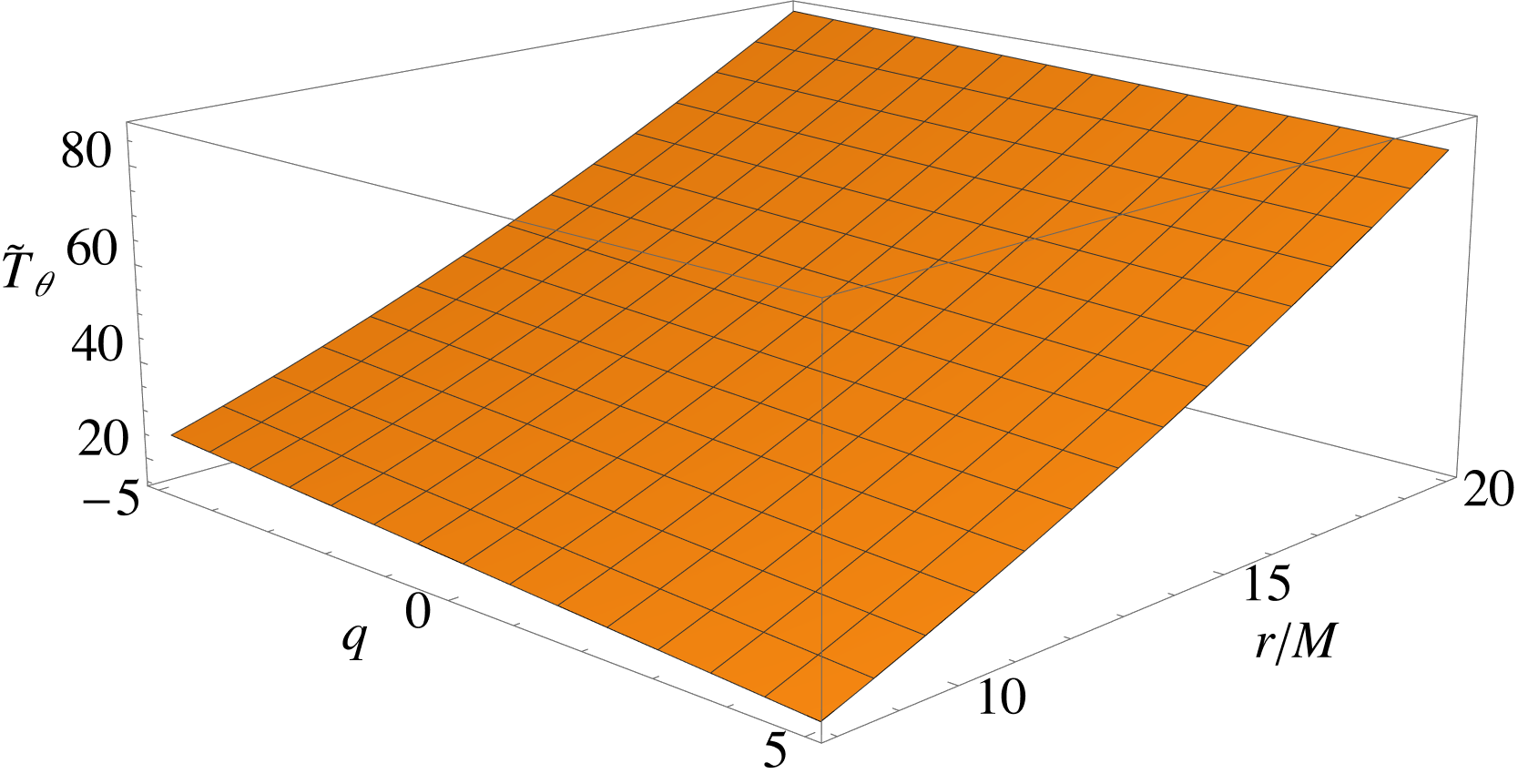}
\hfill} 
{\hfill
\includegraphics[width=8cm]{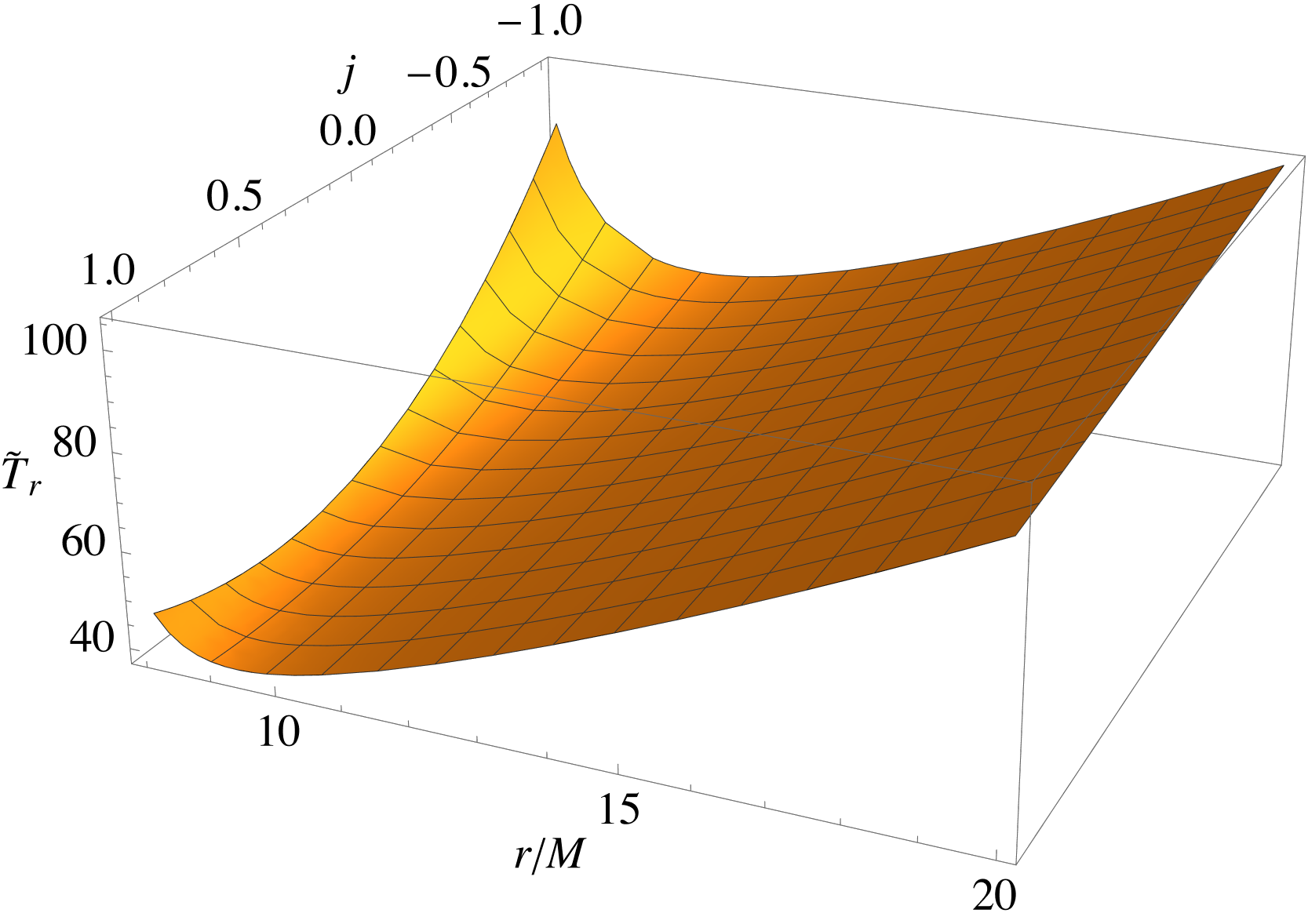}\hfill
\includegraphics[width=8cm]{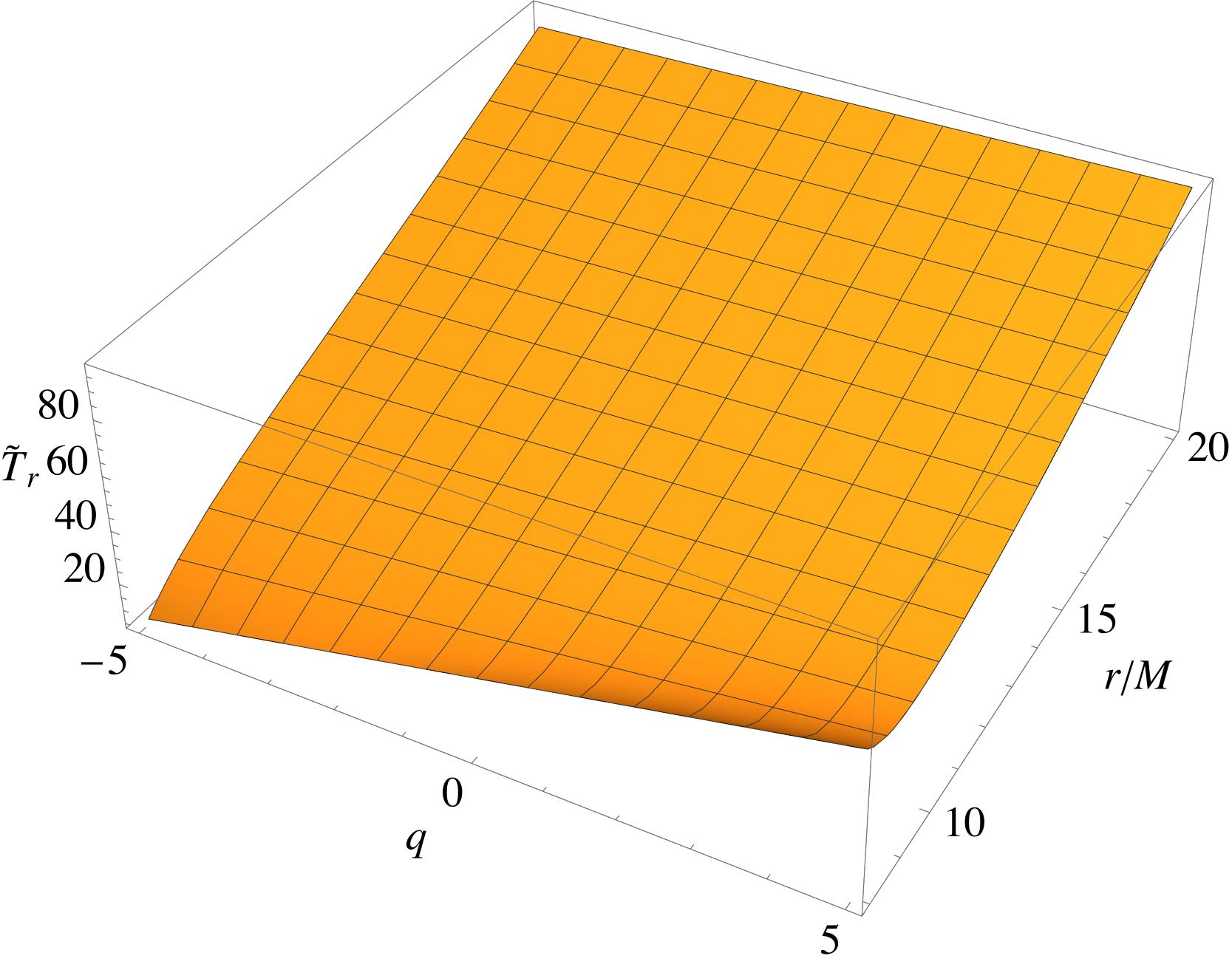}
\hfill} 
\caption{
    Dependence of the dimensionless periods 
    $\tilde{T}_\theta=T_{\theta}/
    M$ and $\tilde{T}_r=T_r/M$ on the parameters of the system in Hartle-Thorne spacetime with fixed $q=2$ (left panel) and $j=0.7$ (right panel). Top left: variation of $\tilde{T}_\theta$ as a function of the angular parameter $j$ and the normalized radius $r/M$. Top right: variation of $\tilde{T}_\theta$ with the deformation parameter $q$ 
    and $r/M$. Bottom left: variation of the radial period $\tilde{T}_r$ as a function of 
    $j$ and $r/M$. Bottom right: dependence of $\tilde{T}_r$ on $q$ and $r/M$. The plots show that both $\tilde{T}_\theta$ and $\tilde{T}_r$ increase 
    with radius, while the dimensionless angular momentum and deformation parameters modify their behavior more significantly in the strong-field region.}
\label{fig:3Dplot} 
\end{figure*}

Both periods grow with radius; increasing the spin $j$ slightly shortens them through frame dragging, while increasing the deformation $q$ towards a more oblate source lengthens them in the strong-field region, all effects fading at large $r$ (Fig.~\ref{fig:3Dplot}).
To obtain the deviation equations for $\xi^3$ and $\xi^4$, we consider the results obtained above in Eqs.~\eqref{eq19} and \eqref{eq20} and substitute the values of $b$ and $c$ in Eqs.~\eqref{eq4} and \eqref{eq5}. The resulting expressions can be written in the following form:
\begin{align}
\xi_{0}^{3} &= \biggl[ \frac{2}{r} + \frac{6 M}{r^{2}} 
- \frac{2 J (31 M + 3 r)}{\sqrt{M r^{7}}} \\ \nonumber
&+ \frac{30 J^{2}}{M r^{4}} 
+ \frac{3 Q (13 M + r)}{M r^{4}} \biggr] e^{i\pi/2} \, \xi_{0}^{1}, \\[6pt]
\xi_{0}^{0} &= \biggl[ 2 \sqrt{\frac{M}{r}} 
- \frac{12 J (9 M + r)}{r^{3}} 
+ \frac{49 J^{2}}{\sqrt{M r^{7}}}\\ \nonumber
&+ \frac{Q (118 M + 9 r)}{2\sqrt{M r^{7}}} \biggr] e^{i\pi/2} \, \xi_{0}^{1}.
\end{align}
Substituting the previously obtained results, Eqs.~\eqref{eq19} and \eqref{eq20} can be rewritten as
\begin{align}\label{allvibrations} \nonumber
\xi^{1} & =\xi_{0}^{1} \sin{\omega s}, \nonumber\\
\xi^{2} & =\xi_{0}^{2}\sin{\Omega s}, \nonumber \\
\xi^{3} & =\xi_0^3 \cos{\omega s},\\ 
\xi^{0} & =\xi_0^0\cos{\omega s}. \nonumber
\end{align}
\begin{figure}
    \centering
    \includegraphics[width=1\linewidth]{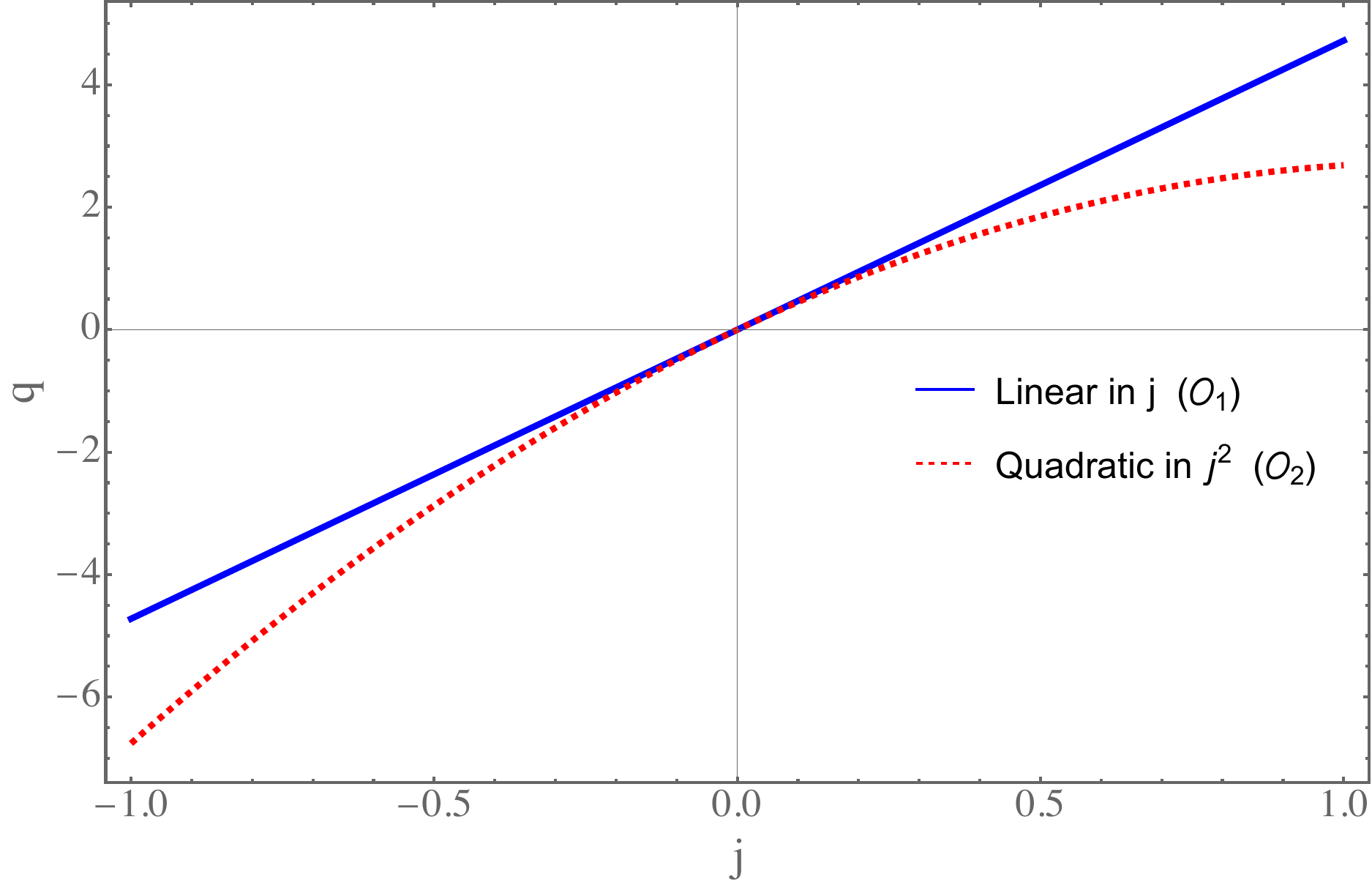}
    \caption{
        Mimicking of the Shirokov effect in the Hartle-Thorne spacetime with $q=Q/M^3$ as a function of the angular momentum $j=J/M^2$ for a neutron star of  $1.4\,M_\odot$. This illustrates how the quadrupole moment can effectively “mimic” the contributions from higher-order mass and angular momentum terms in the vertical displacement. The curves are plotted for a fixed orbital radius $r = 20$ km.}
    \label{fig:mim_Shirokov}
\end{figure}
Based on (\ref{eq:DeltaT}) and (\ref{allvibrations}), if the number of completed oscillations $n= (1,2,3..)$  is relatively small, then the vertical displacement becomes
\begin{align}
\xi^{2}&=\xi_{0}^{2}\sin{\frac{2 \pi n T_{r}}{T_{\theta}}}\nonumber\\
&\approx \xi_{0}^{2}\sin\Bigg[2 \pi n\Bigg(
1
+ \frac{3M}{r}
+ \frac{27M^{2}}{2r^{2}}
\nonumber\\[3pt]
&\quad
- \frac{6J(7M + r)}{\sqrt{M}\,r^{5/2}}
+ \frac{21J^{2}}{M r^{3}}
+ \frac{3Q(19M + 2r)}{2M r^{3}} \Bigg)\Bigg]\\ \nonumber
&\approx
\xi_{0}^{2}\, 2\pi n \Bigg(
\frac{3M}{r}
+ \frac{27M^{2}}{2r^{2}} \\ \nonumber
&\qquad- \frac{6J(7M + r)}{\sqrt{M}\,r^{5/2}}
+ \frac{21J^{2}}{M r^{3}}
+ \frac{3Q(19M + 2r)}{2M r^{3}}
\Bigg).
\label{eq:vertshift}
\end{align}
\begin{figure*}[htbp]
{\hfill
\includegraphics[width=0.495\textwidth]{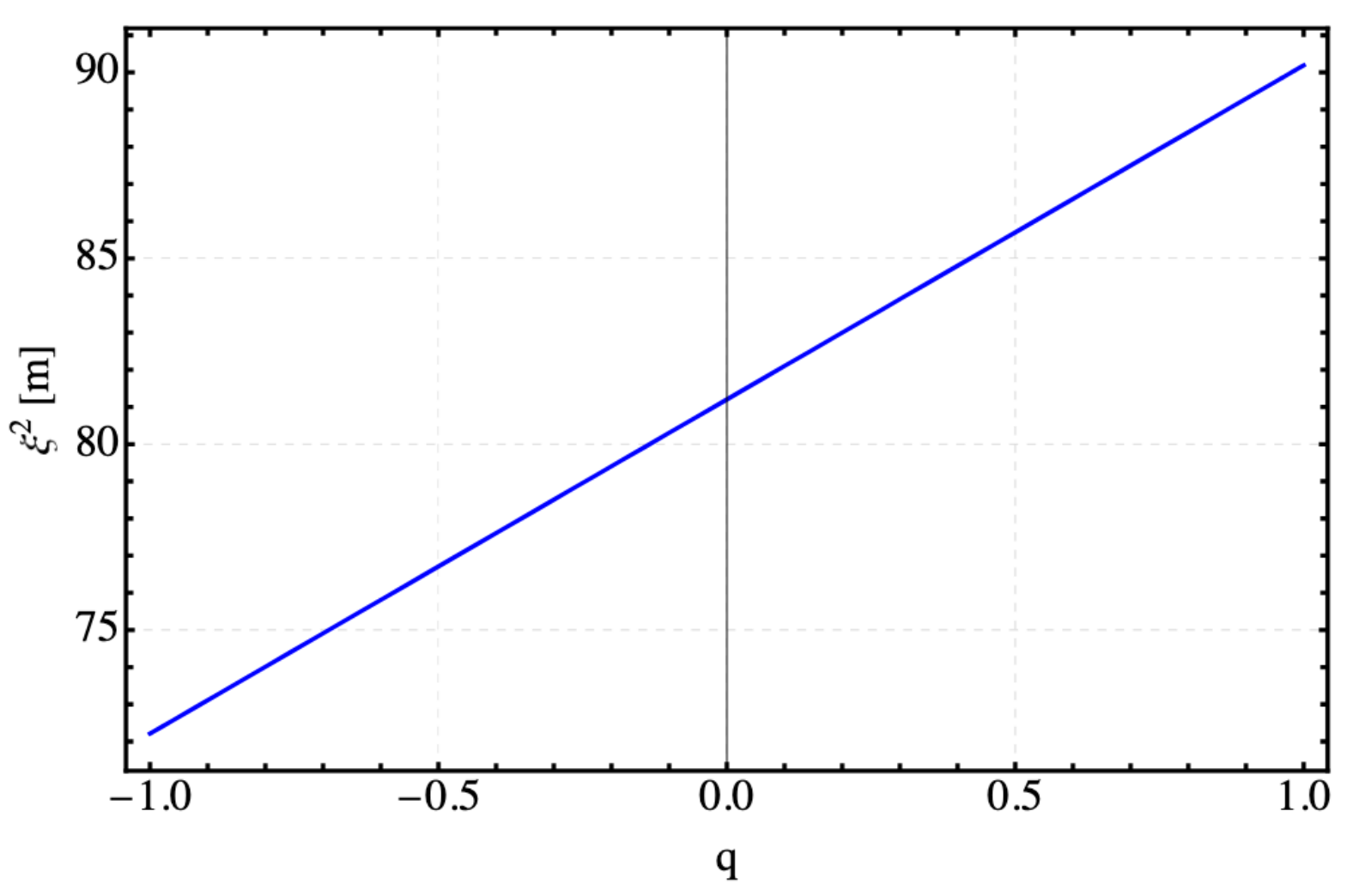}
\includegraphics[width=0.495\textwidth]{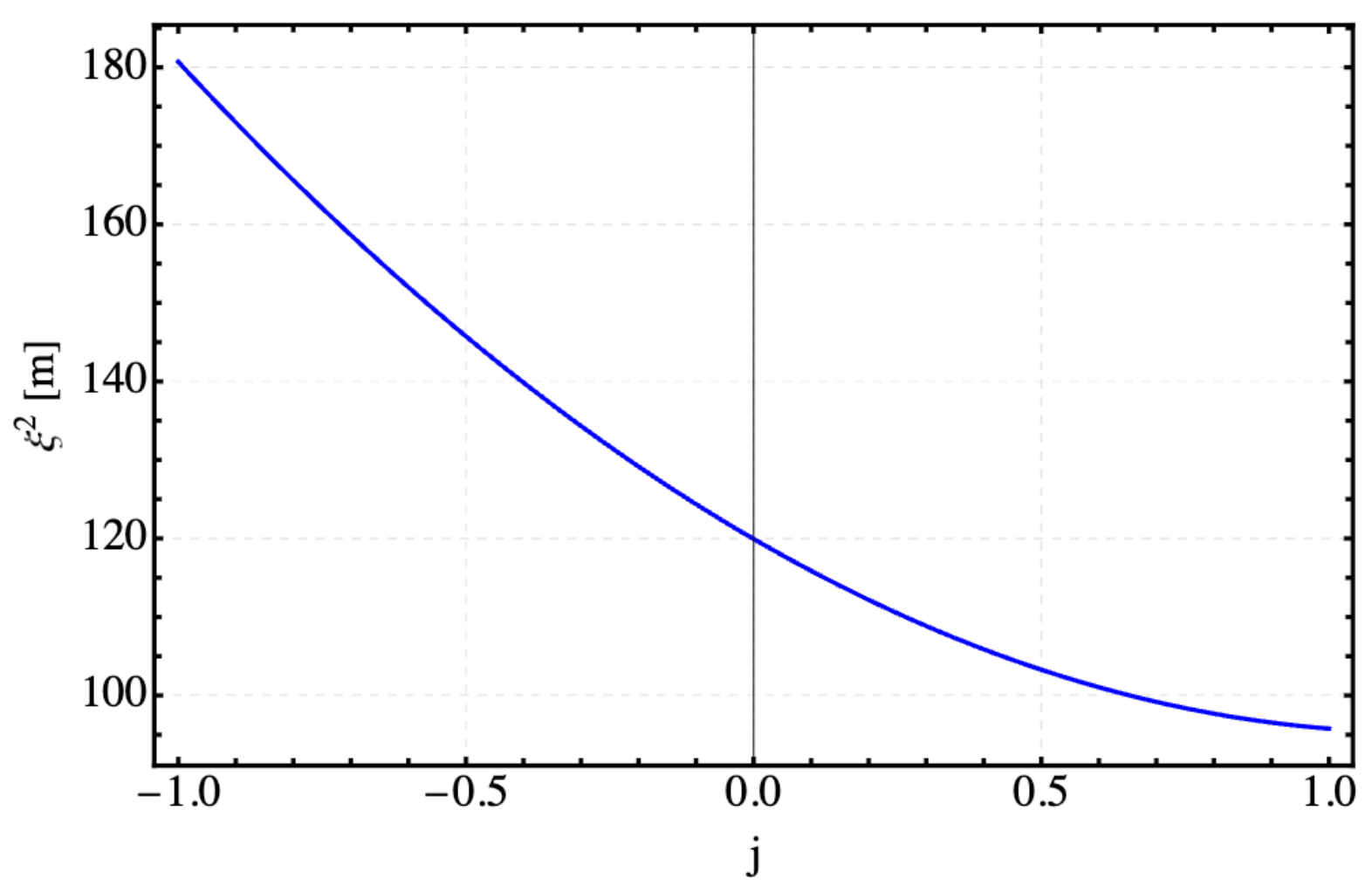}
\hfill}
\caption{Dependence of the vertical displacement for the Shirokov effect in the Hartle-Thorne metric,   $\xi^2$,  on the deformation parameter $q=Q/M^3$ with fixed $j=0.7$ (left panel) and angular momentum $j=J/M^2$ with fixed $q=2$ (right panel). The following values are used:  $\xi_0^2 = 10^{-3}~\text{km}$, $n = 10$, $M=1.4M_{\odot}$ and $r =20~\text{km}$.}
\label{fig:vertdispl}
\end{figure*}
The vertical displacement $\xi^{2}$ increases roughly linearly with the quadrupole parameter $q$ and decreases monotonically with the spin $j$, as shown in Fig.~\ref{fig:vertdispl}.
\subsection{Spin-quadrupole mimicking}\label{sec:shir_mimic}
Based on Eq.\eqref{TroverTtheta}, Figure~\ref{fig:mim_Shirokov}  shows that for a $1.4\,M_\odot$ neutron star  
the quadrupole parameter $q$ inferred from the Shirokov effect depends 
almost linearly on the spin parameter $j$ in the slow-rotation regime. 
This behavior arises because the frame-dragging term (linear in $j$) 
partially compensates the quadrupolar contribution to the vertical motion. 
As a result, different combinations of $(q,j)$ produce nearly the same 
vertical displacement when $|j|\ll 1$. This leads to a spin–quadrupole mimicking effect: a slowly rotating neutron star 
with an appropriate quadrupole moment can reproduce the signal of a differently deformed or differently rotating object. Therefore, the Shirokov effect alone 
cannot uniquely separate spin and quadrupole contributions. Including $j^2$ terms introduces a slight curvature in the 
$q(j)$ relation and formally breaks the exact degeneracy. However, 
for moderate values of $|j|$, this correction remains small and the 
degeneracy is still significant.
\subsection{Static limit: comparison with the $q$-metric}\label{sec:shir_qmetric}

In order to establish a consistent comparison between the quadrupolar structure of the Hartle-Thorne spacetime and that of the $q-$metric, we consider the static limit of the Hartle-Thorne solution by setting the rotation parameter $j$ to zero. This allows us to isolate the purely quadrupolar contribution to the gravitational potential and to directly compare it with the deformation encoded in the parameter $q_{ZV}$ of the $q-$metric \cite{Utepova2025}.  

The comparison is meaningful because both metrics describe axially symmetric, deformed configurations that reduce to the Schwarzschild solution when the deformation vanishes, and because the Hartle-Thorne metric supplies a physically defined mass quadrupole moment $Q$ against which the deformation parameter $q_{ZV}$ of the $q$-metric can be interpreted.  

Following the weak-field expansion of the $q$-metric introduced in~\cite{2019PhRvD..99d4005A} and applied for interpretation purposes in~\cite{2025EPJC...85..319I}, we define the quasi-spherical coordinates $(\rho,\vartheta)$, related to the standard spherical coordinates $(r,\theta)$ by
\begin{align}
r &= \rho \left[ 1 - q_{ZV} \frac{m}{\rho} - q_{ZV} \frac{m^2}{\rho^2}\left( 1 + \frac{m}{\rho}... \right) \sin^2 \vartheta \right], \\
\theta &= \vartheta - q_{ZV} \frac{m^2}{\rho^2}\left( 1 + 2\frac{m}{\rho}... \right)\sin \vartheta \cos \vartheta.
\end{align}

These transformations allow us to express the $q-$metric in the form of a weak-field post-Newtonian expansion. To first order in $q_{ZV}$ and $1/\rho^3$, the line element of the $q-$metric becomes
\begin{align}
ds^2 = (1 + 2\Phi_N) \, dt^2 & - (1 + 2\Phi_N)^{-1} d\rho^2 \nonumber\\
&- U(\rho, \vartheta)\, \rho^2 (d\vartheta^2 + \sin^2\vartheta\, d\phi^2),
\end{align}
where
\begin{align}
\Phi_N &= -\frac{M}{\rho} + \frac{Q}{\rho^3} P_2(\cos\vartheta), \\
U(\rho, \vartheta) &= 1 - 2\frac{Q}{\rho^3} P_2(\cos\vartheta),
\end{align}
and $P_2(\cos\vartheta) = (3\cos^2\vartheta - 1)/2$ is the Legendre polynomial of degree~2.  
The parameters $M$ and $Q$ correspond respectively to the total mass and the quadrupole moment of the source, given by
\begin{equation} \label{M_Q_transf}
M = m(1 + q_{ZV}), \qquad Q = -\frac{2}{3} m^3 q_{ZV}.
\end{equation}

Thus, the deformation parameter $q_{ZV}$ in the $q-$metric and the dimensionless quadrupole moment $q=Q/M^3$ in the Hartle-Thorne spacetime describe the same physical effect--the quadrupolar deviation from spherical symmetry--but differ by a fixed proportionality factor and a sign convention. {The negative sign reflects the convention that a prolate source corresponds to  $q_{ZV} > 0$ (and $Q<0$), whereas an oblate configuration has $q_{ZV} < 0$ (and $Q>0$)}.  

The comparison is performed at fixed multipole content: we match the total mass
$M$ and the mass quadrupole moment $Q$ through Eq.~\eqref{M_Q_transf}, and we
compare the frequencies at equal circumferential radius (and, where applicable, equal orbital frequency). With this prescription, the agreement of the expansions
to first order in $q_{ZV}$ and second order in $M$ is an invariant statement
about two exterior geometries sharing the same $M$ and $Q$, rather than an
artifact of coordinates. Beyond this order the two spacetimes necessarily
differ, since the Hartle-Thorne metric carries an independent angular momentum
while the $q$-metric is static; the comparison is therefore meaningful only as a
controlled cross-check at the stated multipole order.

We distinguish this multipole matching check from the physical degeneracy
emphasized in this work. The latter is the spin--quadrupole ($j$--$q$)
mimicking \emph{within} the Hartle-Thorne family: the leading rotational and
quadrupolar terms enter the invariant ratio and the secular displacement
$\xi^{2}(n)$ in the same combination, so that for $|j|\ll 1$ a range of $(j,q)$
values yields nearly identical signatures. This degeneracy is broken formally at
$\mathcal{O}(j^2)$ but remains observationally significant in the slow-rotation
regime.

Setting $j = 0$ and applying the coordinate transformations above, the $q-$metric reproduces the same post-Newtonian structure as the static limit of the Hartle-Thorne metric. One can then reproduce \cite{Utepova2025} and obtain the fundamental frequencies $\Omega,\omega$ in $(\rho,\vartheta)$ coordinates:
\begin{align}\label{freq_expansions}
\Omega & \approx \sqrt{\frac{m}{\rho^{3}}}
\left[
1 + \frac{3m}{2\rho} + \frac{27 m^2}{8\rho^2}
+ \frac{q_{ZV}}{2}\!\left(1 + \frac{9m}{2\rho} + \frac{159 m^2}{8\rho^2}\right)
\right],\\[4pt]
\omega & \approx \sqrt{\frac{m}{\rho^{3}}}
\left[
1 - \frac{3m}{2\rho} - \frac{45 m^2}{8\rho^2}
+ \frac{q_{ZV}}{2}\!\left(1 - \frac{9m}{2\rho} - \frac{233 m^2}{8\rho^2}\right)
\right].
\end{align}
Applying the mass and quadrupole transformations~\eqref{M_Q_transf} to the Hartle-Thorne frequencies~\eqref{eq:Omega_HT}--\eqref{eq:omega_HT} and expanding to first order in $q_{ZV}$ and second order in $M$, we recover exactly the $q$-metric frequencies above.

\subsection{Slow-rotation limit: comparison with the Lense-Thirring metric}\label{sec:shir_LT}

In \cite{nduka1977shirokov},   the Shirokov effect was evaluated to the first order in the angular momentum parameter \(a/r\) for the Kerr metric in Boyer–Lindquist coordinates, considering circular geodesics at \(r = \mathrm{const}\) in the equatorial plane \(\theta = \pi/2\). However, the analysis does not extend beyond the Lense–Thirring approximation. To facilitate a direct comparison with the present results, 
we reproduce those calculations here, 
{correcting several inaccuracies and typos, and completing the derivation by explicitly obtaining the approximate frequencies \(\Omega\) and \(\omega\) as series expansions in the mass \(M\), retaining terms linear in the spin parameter \(a\)}. We obtain
\begin{align}
\Omega &\approx 
\sqrt{\frac{M}{r^3}}
\Biggl[
1 + \frac{3M}{2r} + \frac{27M^{2}}{8r^{2}} 
\nonumber \\[3pt]
&\hspace{3.5em}
- \frac{3 J_{LT}\left(63 M^{2} + 20M r + 8r^{2}\right)}{8\sqrt{M}\,r^{7/2}} 
\Biggr],
\\[6pt]
\omega &\approx 
\sqrt{\frac{M}{r^3}}
\Biggl[
1 - \frac{3M}{2r} - \frac{45M^{2}}{8r^{2}}
\nonumber \\[3pt]
&\hspace{3.5em}
+ \frac{3J_{LT}\left(231M^{2} + 44M r + 8r^{2}\right)}{8\sqrt{M}\,r^{7/2}}
\Biggr],
\label{eq:Omega_omega_spin}
\end{align}
where $J_{LT}=Ma$ is the angular momentum of  the Lense-Thirring metric.
As can be seen, these expressions are in complete 
agreement with \(\Omega\) ~\eqref{eq:Omega_HT} and \(\omega\) ~\eqref{eq:omega_HT} in the limiting case $J\neq 0$, \(Q = 0\) and \(J^{2} = 0\). 

\subsection{Weak-field limit: dominance of the Newtonian quadrupole}\label{sec:shir_weakfield}
Evaluating the period splitting~\eqref{eq:DeltaT} in the weak-field regime --
restoring the gravitational constant $G$ and the speed of light $c$ and keeping
the leading mass, quadrupole, and spin terms -- the relative difference between
the vertical and in-plane oscillation periods becomes
\begin{equation}\label{eq:DeltaT_weak}
    \frac{\Delta T}{T_0}
    \approx
-\frac{3 Q}{M r^2}-\frac{3 G M}{c^2 r}-\frac{69 G Q}{4 c^2 r^3}+\frac{6 \sqrt{G}J}{c^2 \sqrt{M}}\left(\frac{1}{r}\right)^{3/2},
\end{equation}
where the four contributions are, respectively, the Newtonian quadrupole term
($Q_{1}$), the relativistic (Schwarzschild) mass term ($M$), the
post-Newtonian quadrupole correction ($Q_{2}$), and the relativistic
frame-dragging term ($J$). We stress that the first term carries no factor of
$G/c^{2}$: it is a purely Newtonian effect, arising from the oblateness of the
source through the gradient of its quadrupolar potential, with no analogue in
the spherically symmetric problem Shirokov analysed.

To assess the relative weight of these contributions in Shirokov's own
configuration -- a small satellite on a circular orbit around the Earth -- we
adopt the terrestrial values~\cite{Yoder1995}
$M_{\oplus}=6\times10^{24}\,\mathrm{kg}$,
$Q_{\oplus}=2.66\times10^{35}\,\mathrm{kg\,m^{2}}$,
$J_{\oplus}=6\times10^{33}\,\mathrm{kg\,m^{2}/s}$,
and an orbital radius $r=7\times10^{6}\,\mathrm{m}$. The individual terms then
evaluate to
\begin{eqnarray}
    \frac{\Delta T}{T_0}\left(Q_1\right)&=&-3.06\times10^{-3}\, ,\\
    \frac{\Delta T}{T_0}(M)&=&-2.14\times10^{-9} \, ,\\
    \frac{\Delta T}{T_0}(Q_2)&=&-1.25 \times10^{-11}  \, \\
       \frac{\Delta T}{T_0}(J) &=& 8.45 \times10^{-11} \, .
\end{eqnarray}
The hierarchy of these numbers is the central outcome of this subsection. The
Newtonian quadrupole term $Q_{1}$ exceeds the relativistic mass term $M$ by
about six orders of magnitude, and the relativistic quadrupole and
frame-dragging terms by a further two. Thus, for any realistically oblate body,
the period splitting is governed not by the relativistic curvature effect
Shirokov identified but by the source's \emph{Newtonian} quadrupole moment, the
relativistic contribution surviving only as a small correction. This
reinterpretation is developed in a companion paper~\cite{idrissov2026newtonian},
where the splitting is obtained directly from the Newtonian mass multipoles; the
relativistic corrections in Eq.~\eqref{eq:DeltaT_weak} are precisely the
higher-order $G/c^{2}$ terms that supplement it.

\section{Shapiro time delay} \label{sec:shap}

The time delay of electromagnetic signals propagating near a massive object -- known as the \emph{Shapiro delay} -- constitutes the fourth classical test of general relativity~\cite{shapiro1964fourth}. In his seminal work, Shapiro analyzed radar signals propagating through the Schwarzschild spacetime of the Sun. In the present work, we extend this framework to the strong-field environment of a neutron star, whose exterior spacetime is described by the Hartle-Thorne metric, thereby capturing the combined influence of mass, rotation, and quadrupole deformation on the light-travel time. 

We treat this as a controlled, analytic computation of a Shapiro time delay
in the Hartle-Thorne exterior -- a strong-field analogue of the classical
Solar-System experiment -- rather than as a direct observational prediction.
Light-propagation and timing effects near compact objects have been studied
extensively in the context of pulsar timing, binary pulsars, X-ray burst
oscillations, and hot-spot/pulse-profile modelling
\cite{2016ApJ...820..139C,2014ApJ...792...87P, 2014ApJ...787..136P,2007ApJ...663.1244M,2016ApJ...818..121P}.
In that language, the coordinate delay computed below for a signal propagating
between two bodies $A$ and $B$ orbiting the star, as in the scheme  Fig.~\ref{fig:scheme},
corresponds to a timing residual, which we make precise and compare with existing
quadrupole-order formulae in Sec.~\ref{HTpulsar}.
\subsection{Null geodesics and the time-delay integrand}\label{sec:shap_setup}

Restricting the analysis to the equatorial plane ($\theta = \pi/2$), we consider null geodesics representing photon trajectories. Using the Euler-Lagrange formalism, we obtain the conserved quantities associated with the spacetime symmetries -- the photon energy $E$, angular momentum $L$, and the four-velocity components $\dot{t}$ and $\dot{\phi}$ -- which, together with the null condition \( g_{\mu \nu} u^\mu u^\nu = 0 \), allow us to compute the coordinate of propagation time between emission and reception points. Integration along the photon path then yields the total round-trip Shapiro delay with respect to flat spacetime.

For massless particles one can write the normalization of the four velocity in terms of metric tensor components of \eqref{ht1}
\begin{eqnarray} \label{eqnormShapiro}
g_{tt}\,\dot{t}^{2}
+ g_{rr}\,\dot{r}^{2}
+ g_{\theta\theta}\,\dot{\theta}^{2}
+ g_{\phi\phi}\,\dot{\phi}^{2}
+ 2 g_{t\phi}\,\dot{t}\dot{\phi}
= 0.
\end{eqnarray}
The metric coefficients are independent of the $t$ and $\phi$ coordinates, 
leading to the existence of the conserved specific energy at infinity, $E$, 
and the conserved $z$-component of the specific angular momentum at infinity, $L_z$. 
The $t$- and $\phi$-components of the 4-velocity of a test-particle can thus be written as ~\cite{2011CQGra..28p5001H,2016EL....11630006B}
\begin{eqnarray} \label{eqtdotphidot}
\dot{t} = \frac{E g_{\phi\phi} + L_{z} g_{t\phi}}{g_{t\phi}^{2} - g_{tt} g_{\phi\phi}},
\qquad
\dot{\phi} = - \frac{E g_{t\phi} + L_{z} g_{tt}}{g_{t\phi}^{2} - g_{tt} g_{\phi\phi}}. 
\end{eqnarray}
Then, 
\begin{equation}
 \label{eq:velocity}
\dot{t}=t_0 \left[1 + jW_1(r)+j^2 W_2(r)+qW_3(r)\right],
\end{equation}
where $j=J/M^2$, $q=Q/M^3$,
$t_{0}$ is the time for the Schwarzschild metric, and $W_{1;2;3}$ are: 
\begin{eqnarray}\nonumber
   t_0&:=&\frac{E r}{r-2 M}, \label{eq:t02}\\ \nonumber
    W_1(r)&=&-\frac{2 L M^2}{E r^3}, \label{eq:W1} \\ \nonumber
    W_2(r)&=&-\left[8 M r^4 (r-2 M)\right]^{-1}\Big(16 M^6+24 M^5 r \\ \nonumber
          &-&8 M^4 r^2-10 M^3 r^3-20 M^2 r^4\\ \nonumber 
           &-&15 r^6+45 M r^5\Big)-W(r) ,\label{eq:W2} \\ \nonumber
     W_3(r)&=&-5\left[8M r(r-2M)\right]^{-1}\Big(2 M^3+4 M^2 r \\ \nonumber
         &+&3 r^3-9 M r^2\Big)+W(r),\\
    W(r) &=&\frac{15r(r-2M)}{16M^2}\ln \left(\frac{r}{r-2M}\right).\label{eq:W}\nonumber
\end{eqnarray}
 \begin{equation}
 \label{eq:velocity2}
\dot{\phi}=\phi_0\left[1 + j V_1(r)+j^2 V_2(r)+q V_3(r)\right],
\end{equation}
where $j=J/M^2$, $q=Q/M^3$,
$\phi_{0}$ is  the angular velocity  for the Schwarzschild metric, and $V_{1;2;3}$ are:
\begin{eqnarray}\nonumber
   \phi_0&:=&\frac{L}{r^2}, \label{eq:t0}\\ \nonumber
    V_1(r)&=&\frac{2 E M^2}{L (r-2 M)}, \label{eq:W111} \\ \nonumber
    V_2(r)&=& \left[8 M r^4 (r-2 M)\right]^{-1}\Big(32 M^6-32 M^5 r \\ \nonumber
    &-&8 M^4 r^2-20 M^3 r^3+40 M^2 r^4\\ \nonumber 
    &+&15 M r^5-15 r^6\Big)+V(r), \\ \nonumber
    V_3(r)&=&-\frac{5  \left(2 M^2-3 M r-3 r^2\right)}{8 M r} - V(r),  \\ \nonumber 
     V(r)&=&\frac{15 \left(r^2-2 M^2\right)}{16 M^2}\ln \left(\frac{r}{r-2M}\right).\nonumber
\end{eqnarray}
Furthermore,  to remove the affine parameter $s$, we use the relationship 
\begin{equation} \label{eqrs}
\frac{dr}{d s} = \frac{dr}{dt} \frac{dt}{ds} =  \frac{dr}{dt} \dot{t}.
\end{equation}
At $r = r_C$ , we have $dr/dt= 0$, and combining \eqref{eqtdotphidot} with \eqref{eqnormShapiro}, we compute $L$ \footnote{Since the parameter $L_{0}$ can take both positive and negative values, all subsequent analytical and numerical investigations are restricted to the corotating case $(L
_{0}>0)$, with the remaining terms taken into account.} for the Hartle-Thorne metric and obtain  
\begin{equation}\label{eq:angmomentum}
   L=  L_0\left[1- jH_1(r)+j^2 H_2(r)+q H_3(r)\right],   
\end{equation}
where $L_{0}$ denotes the  angular momentum  for the Schwarzschild metric and the functions $H_{1;2;3}$ are:

\begin{eqnarray}  \nonumber
L_{0}&:=&\frac{r_C E}{\sqrt{1-2M/r_C}}, \label{eq:L0} \\ \nonumber
H_{1}(r)&=&\frac{2 M^2}{r_C^{3/2}(r_C-2M)^{1/2}},  \label{eq:H1}\\ \nonumber
H_{2}(r)&=&-\left[8Mr_C^4(r_C-2 M)\right]^{-1}\Big(24 M^6-20 M^5 r_C  \label{eq:H3}\\ \nonumber
            &-&8 M^4 r_C^2-15 M^3 r_C^3+10 M^2 r_C^4\\ \nonumber
            &+& 30 M r_C^5 -15 r_C^6\Big)+H(r) ,\\  \nonumber
    H_{3}(r)&=&-\left[8 M r_C(r_C-2M)\right]^{-1}\Big(15 M^3  \label{eq:H2} \nonumber \\
            &-&10 M^2 r_C - 30 M r_C^2+15 r_C^3\Big) - H(r),  \nonumber \\
     H(r)&=&\frac{15 \left(M^2+M r_C-r_C^2\right) }{16 M^2} \ln \left(\frac{r_C}{r_C-2 M}\right)  \label{eq:H}\nonumber.
\end{eqnarray}

To compute the Shapiro time delay in the Hartle-Thorne spacetime, we need the integrand  $1/\sqrt{I_1}$ (cf. Eq.(\ref{integral_general} below), which after straightforward but lengthy analytical manipulations can be expressed as 
\begin{equation}\label{eq:integrandgen}
   \frac{1}{\sqrt{I}}= \left(\frac{1}{\sqrt{I}}\right)_0 \left[1+ j \left(\frac{1}{\sqrt{I}}\right)_1 + j^2 \left(\frac{1}{\sqrt{I}}\right)_2+ q \left(\frac{1}{\sqrt{I}} \right)_3 \right].   
\end{equation}
The leading coefficients $(1/\sqrt{I})_{0}$ and $(1/\sqrt{I})_{1}$ are given
below; the second-order spin and quadrupole coefficients $(1/\sqrt{I})_{2}$ and
$(1/\sqrt{I})_{3}$, together with the polynomial $\mathcal{P}(r,r_{C})$, are
collected in Appendix~\ref{app:shapiro_coeffs}.
\begin{align} \nonumber
\left(\frac{1}{\sqrt{I}}\right)_0
&=
\frac{r^{5/2} \sqrt{2M-r_C}}
{(r - 2M)\,
\sqrt{
r^3(2M - r_C) + r_C^3(r-2M)
}}, \nonumber
\end{align}
\begin{align} \nonumber
\left(\frac{1}{\sqrt{I}}\right)_1
&=
\frac{2 M^2 (2M - r)\, r_C^{3/2}\, (r^2 + r r_C + r_C^2)}
{r^3 \sqrt{r_C - 2M}}
\nonumber\\
&\quad \times
\frac{1}{
2M (r^2 + r r_C + r_C^2)
- r r_C (r + r_C).
} \nonumber
\end{align}
In subsection~\ref{HTsection}, we will evaluate in full length the integral of this expression, which determines the time delay in the approximate Hartle-Thorne spacetime. 
But first we examine the  weak-field limits in $\mathcal{O}(J)$ and $\mathcal{O}(M)$, which correspond to a slowly rotating compact object (\ref{LTsection})  and in $\mathcal{O}(Q)$ and $\mathcal{O}(M)$ to a  static quadrupole deformed source (\ref{static}).
\subsection{Slow-rotation limit: the Lense-Thirring effect ($j\neq0$, $q=0$, $j^2\approx0$)}\label{LTsection}
In this subsection, we examine the influence of stellar rotation on the Shapiro time delay by considering the Hartle-Thorne metric in the limit $q = 0$, where the quadrupole deformation is neglected. The remaining parameter $j$ represents the dimensionless angular momentum of the source, which introduces the frame-dragging (Lense-Thirring) effect and leads to an azimuthal asymmetry in photon propagation. By varying $j$ in the range $[-1, 1]$, we investigate how rotation modifies the time delay with respect to the static Schwarzschild configuration. To this end, we replace the expressions  \eqref{eqtdotphidot} and \eqref{eq:angmomentum} in the general equation \eqref{eqnormShapiro}, which allows us to compute the time interval as 
\begin{equation} \label{integral_general}
dt = \pm \frac{1}{c} \frac{dr}{\sqrt{I_1}},
\end{equation} 
where the integrand is given explicitly in Eq.(\ref{eq:integrandgen}), which in the limiting case  $q=0$, $j^2\approx 0$, and to first order in $\mathcal{O}(J)$ and $\mathcal{O}(M)$\footnote{{For convenience, we use  the dimensionless parameters $j = J/M^2$ and $q = Q/M^3$. However, all analytical calculations -- particularly the expansions -- were carried out in terms of the physical parameters $J$ and $Q$, with expressions rewritten in terms of $j$ and $q$ only at the stage of result presentation and numerical evaluation.}}  reduces to 
\begin{align} \label{I1}
\frac{1}{\sqrt{I_1}}&\approx \sqrt{\frac{r^2}{r^2 - r_C^2}} \\  \nonumber
& \times \Bigg[1 + \frac{M (2 r + 3 r_C)}{r (r + r_C)} +\frac{2 j M^2 (r^2 + r r_C + r_C^2)}{r^4 \, r_C \, (r + r_C)^2}
\\
&\qquad\times \Big(
r r_C (r + r_C)
+ 3M (r^2 + r r_C + r_C^2)
\Big)
\Bigg].\nonumber
\end{align}

\begin{figure*}
    \centering
    \includegraphics[width=0.8\linewidth]{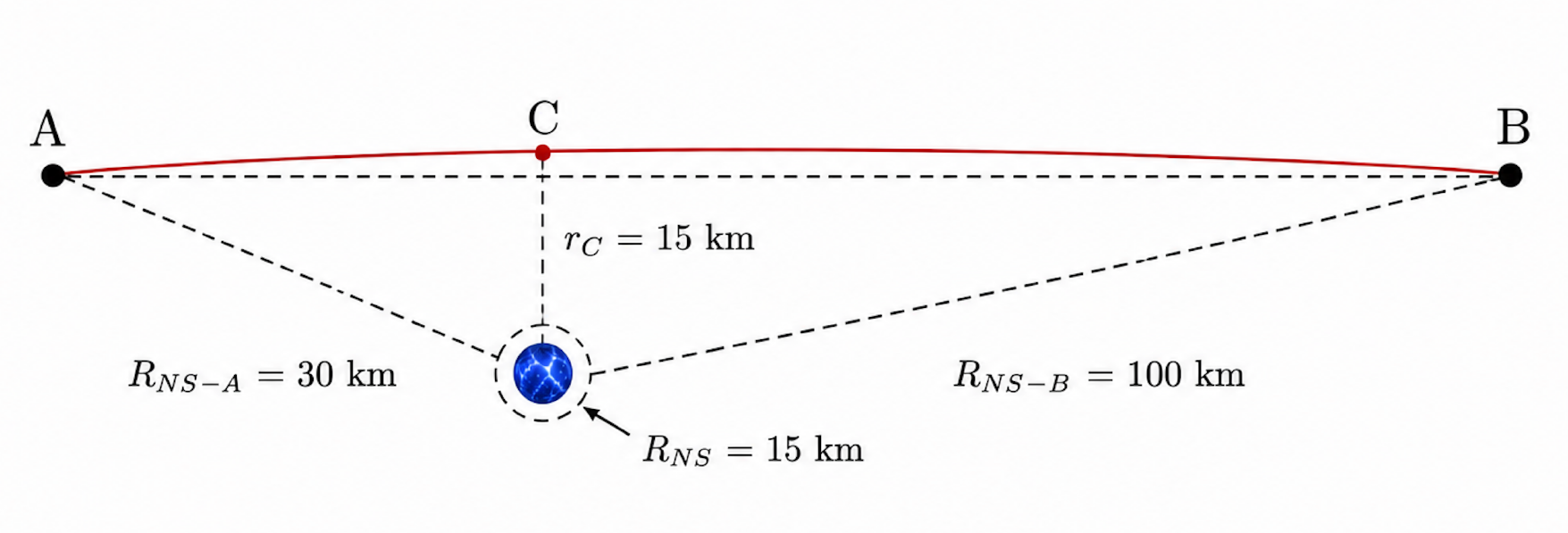}
    \caption{Schematic illustration of the Shapiro time delay. A compact object (e.g., a neutron star) is located at the center. Two objects, A and B, move along circular orbits around the central mass. Point C marks the closest approach to the gravitating source for a ray propagating from A to B, resulting in an additional travel time compared to the flat space case.}
    \label{fig:scheme}
\end{figure*}
\begin{figure*}[htbp]
{\hfill
\includegraphics[width=0.495\linewidth]{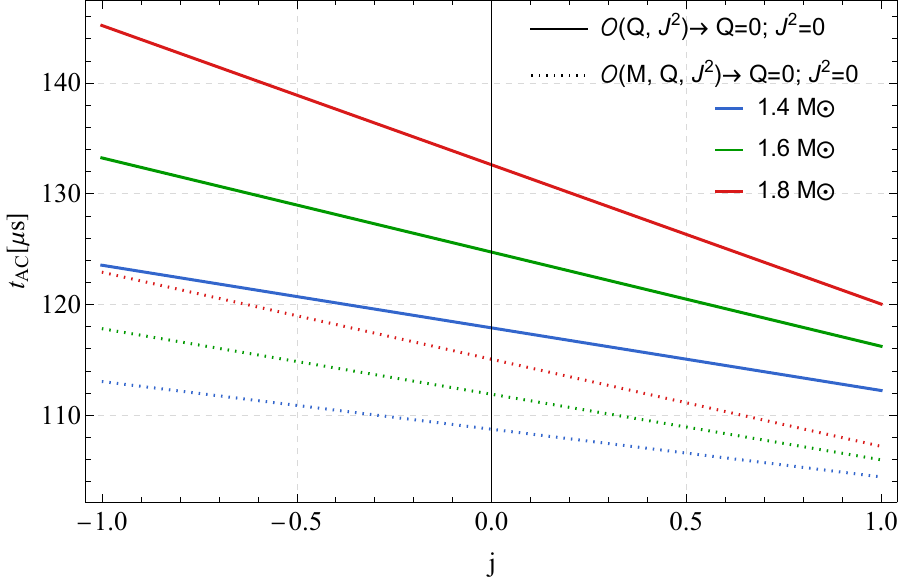}
\includegraphics[width=0.495\linewidth]{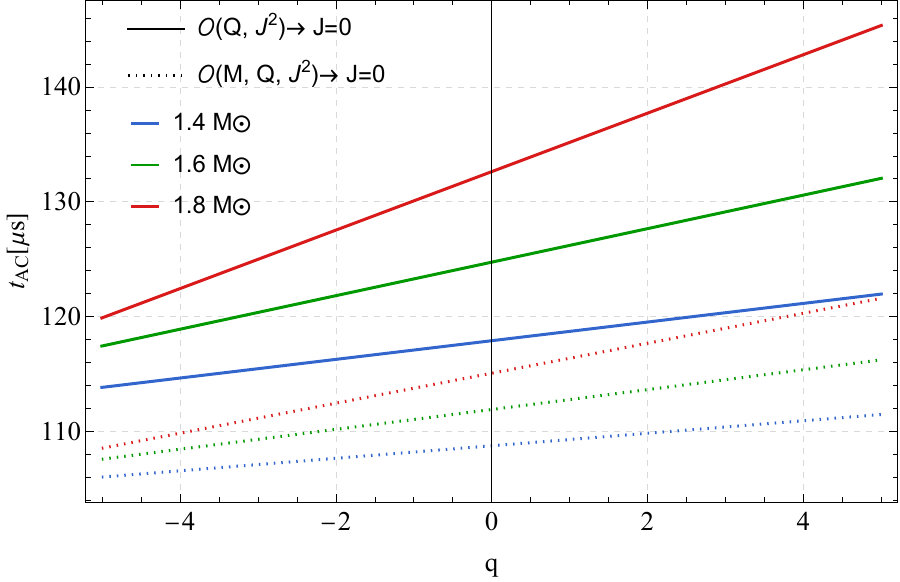}
\hfill} 
\caption{Numerical integration of the time delay as a function of the angular momentum $j=J/M^2$ (left panel), where $q=0$, $j^2=0$ and  for different mass of the compact object $M$ in solar masses and as a function of deformation parameter $q=Q/M^3$ (right panel), where $j=0$. The dashed curves correspond to the expansion up to $\mathcal{O}(M, Q, J^2)$, while the solid lines correspond to $\mathcal{O}(Q, J^2)$; here and further, these expansions are derived from the same Eq.~\eqref{eq:integrandgen}.} 
\label{fig:numinttime}
\end{figure*}
The time it takes for a light ray signal to go from A to C in Fig.~\ref{fig:scheme} is 
\begin{equation} \label{integral_q0}
t_{AC} = \frac{1}{c} \int_{r_C}^{r_A} \frac{dr}{\sqrt{I_1}}.
\end{equation}
\begin{figure*}[htbp]
{\hfill
\includegraphics[width=0.495\linewidth]{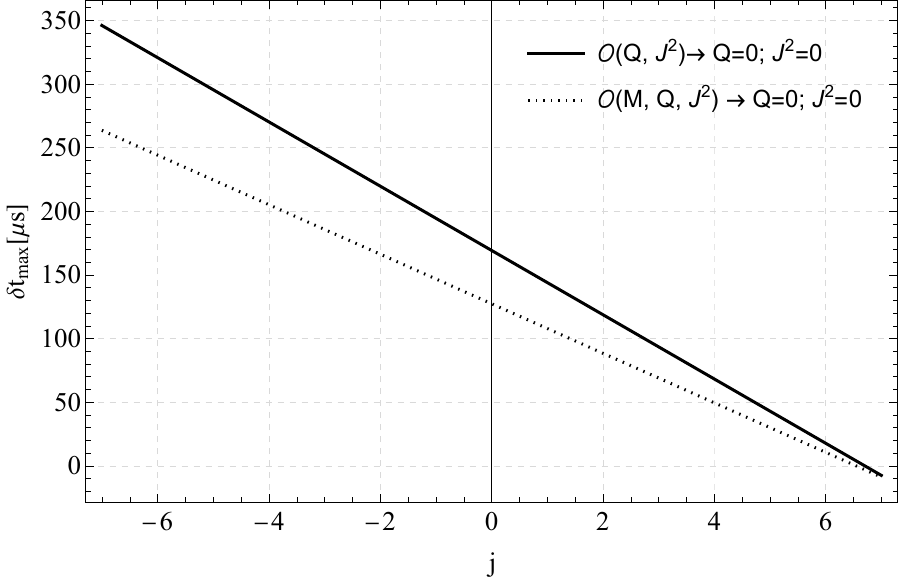}
\includegraphics[width=0.495\linewidth]{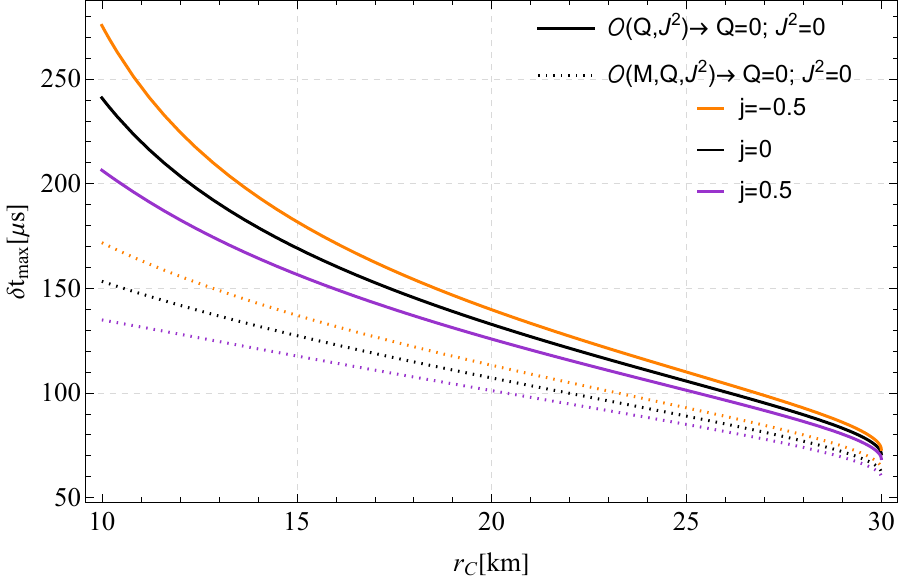}
\hfill} 
\caption{The time delay for Lense-Thirring with respect to the angular momentum  $j=J/M^2$ (left panel) when a radar signal travels
back and forth through the path close to the neutron star, when the exterior field is described by the Hartle-Thorne metric, according to Eq. \eqref{tACvsj} where $q=0$, $j^2=0$ and  $1.4M_\odot$ mass of neutron star. Time delay as a function of the distance from the source of gravity to the propagating light path $r_{C}$ for prograde $j=0.5$ and retrograde $j=-0.5$ rotation of neutron star, $j=0$ corresponds to the Schwarzschild case (right panel).}
\label{fig:td_round_vs_j}
\end{figure*}
The numerical integration of Eq.~\eqref{integral_q0} is shown in the left panel of Fig.~\ref{fig:numinttime}. The one-way delay $t_{AC}$ decreases linearly with $j$: frame dragging shortens the coordinate travel time for prograde photons ($j>0$) and lengthens it for retrograde ones ($j<0$), and the slope steepens with $M$ as stronger fields amplify both curvature and frame dragging.

Eventually, the time that it takes for the electromagnetic signal to go from point A to point C turns out to be
\begin{align} \label{tACvsj} \nonumber
t_{AC} &=
\frac{\sqrt{r_A^2 - r_C^2}}{c r_A^2 r_C^2 (r_A + r_C)^2}
\Big[
r_A^2 r_C^2 (r_A + r_C)(M + r_A + r_C) \\ \nonumber
&\qquad
- j M^2 \Big(
2 r_A r_C (2 r_A^2 + 3 r_A r_C + r_C^2)
\\ \nonumber
&\qquad\qquad
+ M (-8 r_A^3 - 7 r_A^2 r_C + 6 r_A r_C^2 + 3 r_C^3)
\Big)
\Big]
\\ \nonumber
&\quad
+ \frac{30 j M^3}{c r_C^2} \left(
\arctan\!\left(\frac{r_A-\sqrt{r_A^2-r_C^2}}{r_C}\right)
- \frac{\pi}{4}\right)
\\
&\quad
+  \frac{2M}{c} \ln\!\left(
\frac{r_A + \sqrt{r_A^2 - r_C^2}}{r_C}
\right),
\end{align}
where $c$ is the speed of light.
In the case of a flat spacetime, the time that the electromagnetic signal would take to go from point \(A\) to point \(C\) is
\begin{equation}
\tilde{t}_{AC}=\frac{1}{c}\sqrt{r_{A}^{2}-r_{C}^{2}}\,,
\end{equation}
and corresponds to the leading order term in \eqref{tACvsj}. The total time that the electromagnetic signal takes to go from \(A\) to \(B\) and come back to \(A\) is
\begin{equation}
t_{\mathrm{tot}}=2t_{AC}+2t_{BC}\,,
\end{equation}
while in flat spacetime it would be
\begin{equation}
\tilde{t}_{\mathrm{tot}}=2\tilde{t}_{AC}+2\tilde{t}_{BC}=\frac{2}{c}\sqrt{r_{A}^{2}-r_{C}^{2}}+\frac{2}{c}\sqrt{r_{B}^{2}-r_{C}^{2}}\,.
\end{equation}
In a previous work \cite{Utepova2025}, the time-delay expression was derived under the simplifying assumption \( R_\odot \ll r_A, r_B \), appropriate for solar-system configurations where the central source is small compared to the planetary distances. Here the central object is a neutron star, for which this approximation no longer holds; we therefore adopt its actual scale, $R_{\mathrm{NS}} = 15~\mathrm{km}$, and place the source and receiver at $r_A = 30~\mathrm{km}$ and $r_B = 100~\mathrm{km}$. In this near-field configuration the photon samples regions of strong curvature, so the delay depends more strongly on the stellar mass and angular momentum.

The maximum round-trip time delay as a function of $j$ relative to flat spacetime is given by
\begin{align} \label{td_round}
    \delta t_{\mathrm{max}}(j) =  t_{\mathrm{tot}}-\tilde{t}_{\mathrm{tot}}.
\end{align}

Figure~\ref{fig:td_round_vs_j} shows $\delta t_{\max}$ for $r_A = 30~\mathrm{km}$ and $r_B = 100~\mathrm{km}$. The left panel gives its dependence on $j$ at fixed $r_C = 15~\mathrm{km}$: the delay decreases nearly linearly, as first-order frame dragging dominates in the slow-rotation regime, with the sign set by the photon's orientation relative to the spin axis. The right panel gives its dependence on $r_C$ for co-rotating ($j=0.5$) and counter-rotating ($j=-0.5$) cases: $\delta t_{\max}$ decreases with $r_C$ as the ray samples weaker field, and the prograde--retrograde offset (a frame-dragging asymmetry, with the retrograde ray now delayed more) shrinks as the curves converge at large $r_C$.

\subsection{Static limit: deformed source ($j=0$, $q\neq0$)}\label{static}

To isolate the role of the angular momentum, we next analyze the case of a compact static but deformed object by setting $j = 0$ and retaining a nonvanishing deformation parameter $q$. This configuration corresponds to an axisymmetric, deformed source without rotation, allowing us to quantify the impact of purely quadrupolar corrections on the Shapiro time delay. The comparison between the rotating and static deformed cases highlights the distinct contributions of frame dragging and geometric deformation to the photon propagation time.

If we consider only up to $\mathcal{O}(M,Q)$ terms in the expansion, we can get
\begin{equation} \label{integral_general2}
dt = \pm \frac{1}{c} \frac{dr}{\sqrt{I_2}},
\end{equation}
where
\begin{align} \label{I2} \nonumber
\frac{1}{\sqrt{I_2}}
&\approx 
\sqrt{\frac{r^2}{r^2 - r_C^2}}\\ 
&\times \Bigg[ 1 
+ \frac{M (2r + 3r_C)}{r (r + r_C)} \\ \nonumber
&\quad +\frac{q M^3}{4 r^4 r_C^2 (r + r_C)^2}
\\ \nonumber
&\quad\times \Big[
4 r r_C \left(r^3 + 3 r^2 r_C + 4 r r_C^2 + 2 r_C^3 \right)
\\ \nonumber
&\quad + M \left(
19 r^4 + 38 r^3 r_C + 70 r^2 r_C^2
+ 86 r r_C^3 + 47 r_C^4
\right)
\Big]
\Bigg]. \nonumber
\end{align}
\begin{figure*}[htbp]
{\hfill
\includegraphics[width=0.495\linewidth]{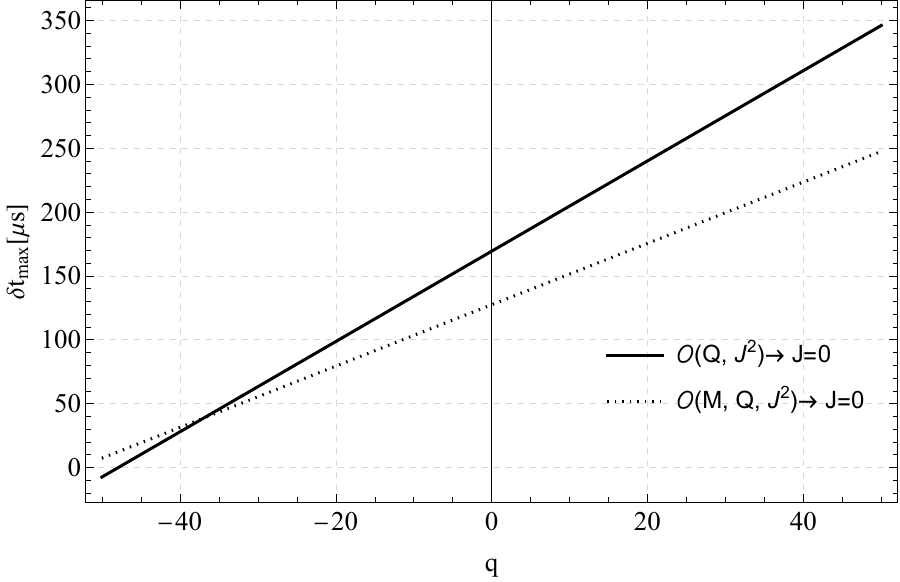}
\includegraphics[width=0.495\linewidth]{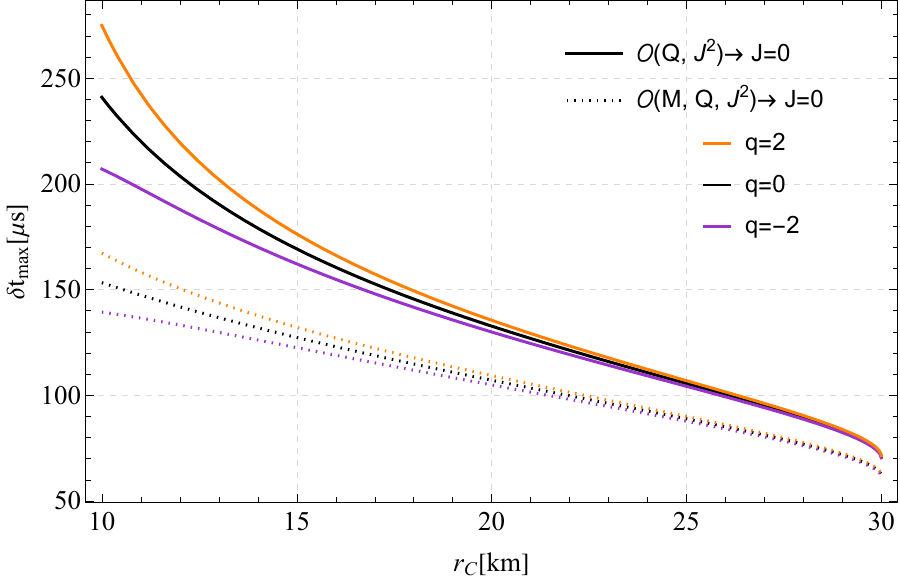}
\hfill} 
\caption{The time delay for static-deformed source with respect to the deformation parameter  $q=Q/M^3$ (left panel) when a radar signal travels
back and forth through the path close to the neutron star, the exterior field is described by the Hartle-Thorne metric, according to Eq. \eqref{tACvsq} where $j=0$, $j^2=0$ and the mass of the neutron star  is $1.4M_\odot$ .  Time delay as a function of the distance from source of gravity to the propagating light path $r_{C}$ (right panel) for an oblate source with $q=2$ and a prolate source with $q=-2$. The Schwarzschild case corresponds to $q=0$.}
\label{fig:td_round_vs_q}
\end{figure*}
According to Fig. \ref{fig:scheme}, the time it takes for a light ray signal to travel from A to C is 
\begin{equation} \label{integral_j0}
t_{AC} = \frac{1}{c} \int_{r_C}^{r_A} \frac{dr}{\sqrt{I_2}}.
\end{equation}

The right panel of Fig.~\ref{fig:numinttime} shows $t_{AC}$ versus $q$ at $j=0$: the delay increases with $q$, since oblate configurations ($q>0$) lengthen the effective path through higher-order curvature corrections. The dependence on $q$ is, however, weaker than on $j$ -- the same parameter range produces smaller deviations -- confirming that in slowly rotating objects frame dragging is the dominant correction and quadrupole deformation a secondary one.

Eventually, the time that it takes for the electromagnetic signal to go from point A to point C turns out to be

\begin{align} \label{tACvsq}
t_{AC} &=
\frac{\sqrt{r_A^2 - r_C^2}}{c r_A^2 r_C^3 (r_A + r_C)^2}
\Big[
M r_A^2 r_C^3 (r_A + r_C) \\ \nonumber
& \qquad+ r_A^2 r_C^3 (r_A + r_C)^2
\\ \nonumber
&\qquad
+ M^3 q\, r_A r_C (3 r_A^2 + 5 r_A r_C + 2 r_C^2)
\\ \nonumber
&\qquad
+ \frac{M^4 q}{8} \left(
-64 r_A^3 - 41 r_A^2 r_C + 78 r_A r_C^2 + 47 r_C^3
\right)
\Big]
\\ \nonumber
&
- \frac{125 M^4 q}{4 c r_C^3}
\arctan\!\left(\frac{r_A-\sqrt{r_A^2 - r_C^2}}{r_C}\right)+\frac{125 \pi  M^4 q}{16 c r_C^3}\\ \nonumber
& + \frac{2 M}{c} \ln\!\left(
\frac{r_A + \sqrt{r_A^2 - r_C^2}}{r_C}
\right),
\end{align}
where $c$ is the speed of light.
In close analogy with the Lense–Thirring case discussed in ~\ref{LTsection}, we apply the same algorithmic procedure, yielding the maximum round-trip time delay as a function of $q$ with respect to flat spacetime, evaluated at \( r_{C} = R_{\mathrm{NS}} \):
\begin{align} \label{tdq_round}
\delta t_{\mathrm{max}} (q)  &=t_{\mathrm{tot}}-\tilde{t}_{\mathrm{tot}}.
\end{align}
Figure~\ref{fig:td_round_vs_q} presents $\delta t_{\max}$ at fixed $j=0$, with $r_A = 30~\mathrm{km}$ and $r_B = 100~\mathrm{km}$. The left panel ($\delta t$ versus $q$ at $r_C = R_{\mathrm{NS}} = 15~\mathrm{km}$) shows a monotonic increase with $q$: an oblate source ($q>0$) redistributes mass toward the equator, deepening the equatorial potential well sampled by the photon and increasing the integrated delay. The right panel ($\delta t$ versus $r_C$ for prolate $q=-2$, Schwarzschild $q=0$, and oblate $q=2$) shows a monotonic decrease with $r_C$, with the curves systematically offset -- oblate above, prolate below the spherical case. The quadrupole-driven offset is comparable to that from moderate changes in $r_C$, so stellar deformation constitutes a significant systematic uncertainty in timing-based radius estimates.

In summary, these results quantify how the Shapiro delay serves as a simultaneous probe of both the mass-radius relation and the quadrupole moment of a massive object. In principle, measurements of this effect could contribute to constraining deviations from spherical symmetry and provide insights into the interior structure of neutron stars.
\subsection{Strong-field regime: full Hartle-Thorne case ($j\neq0$, $j^2\neq0$, $q\neq0$)}\label{HTsection}

\begin{figure*}[htbp]
{\hfill
\includegraphics[width=0.495\linewidth]{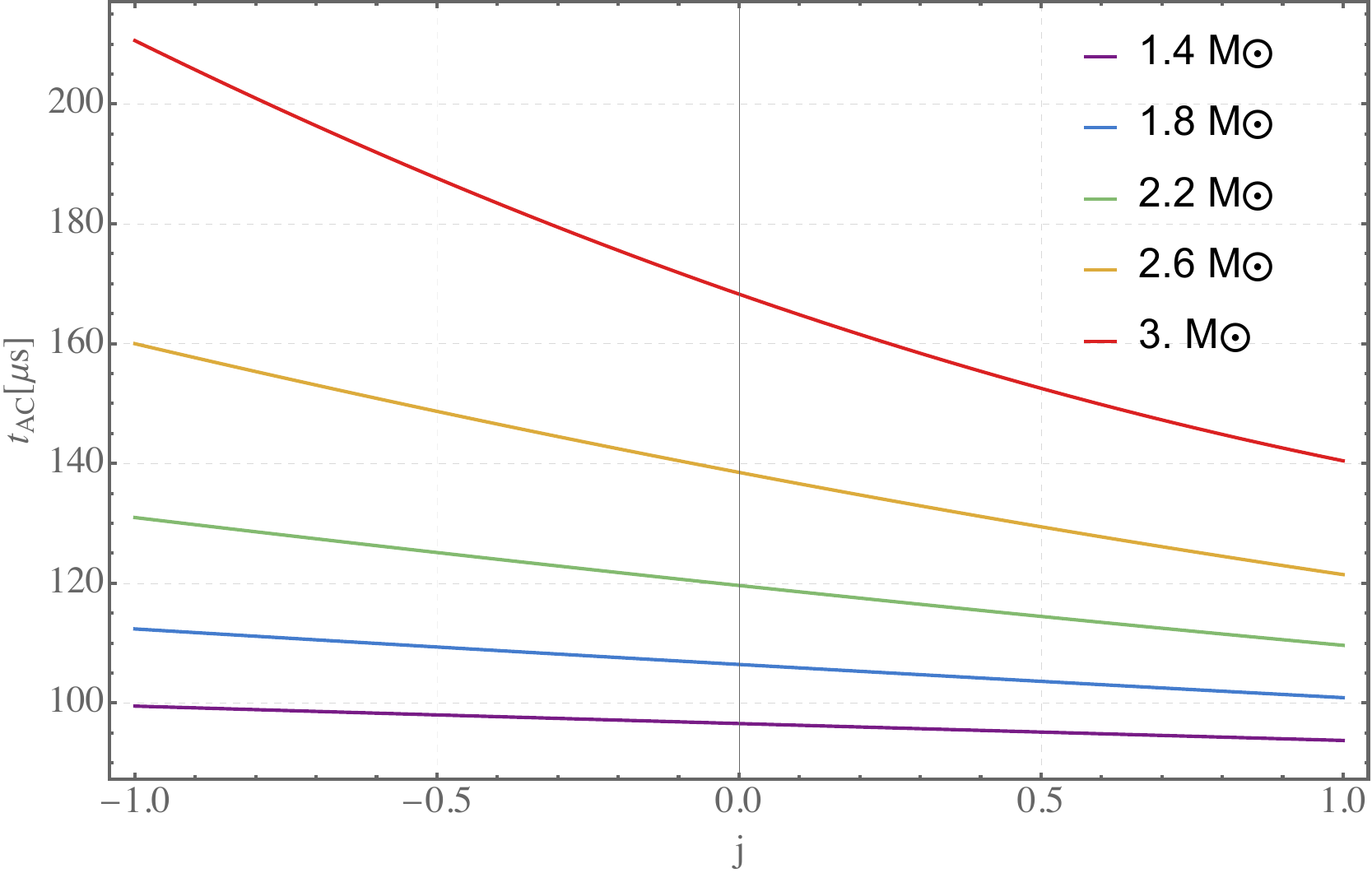}
\includegraphics[width=0.495\linewidth]{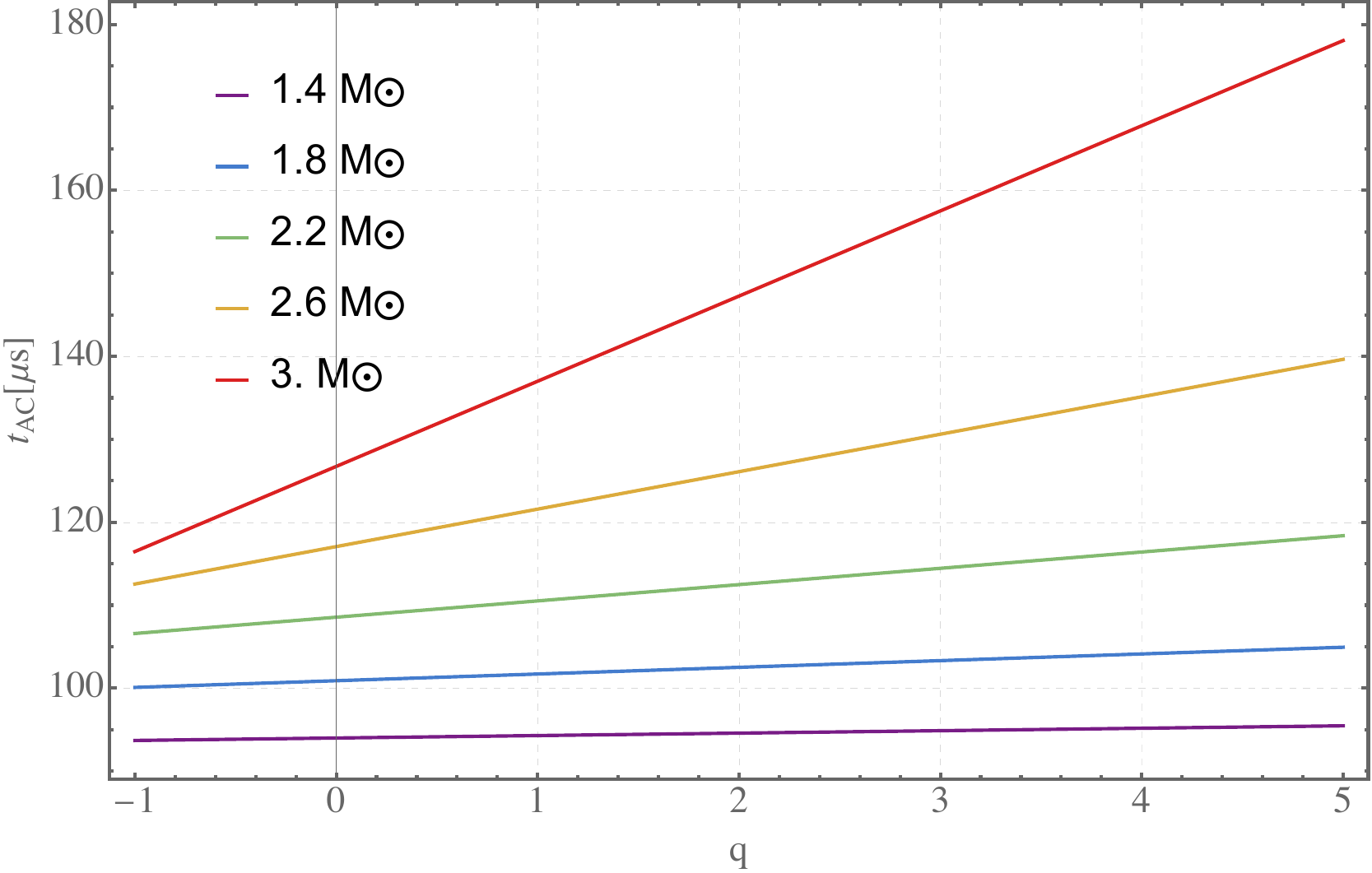}
\hfill} 
\caption{Numerical integration of time delay for Hartle-Thorne as a function of angular momentum $j=J/M^2$ (left panel) with fixed $q=2$  for different mass of the compact object $M$ in solar masses and as a function of deformation parameter $q=Q/M^3$ with fixed $j=-0.7$ (right panel).}
\label{fig:numinttime_HT}
\end{figure*}
\begin{figure*}[htbp]
{\hfill
\includegraphics[width=0.495\linewidth]{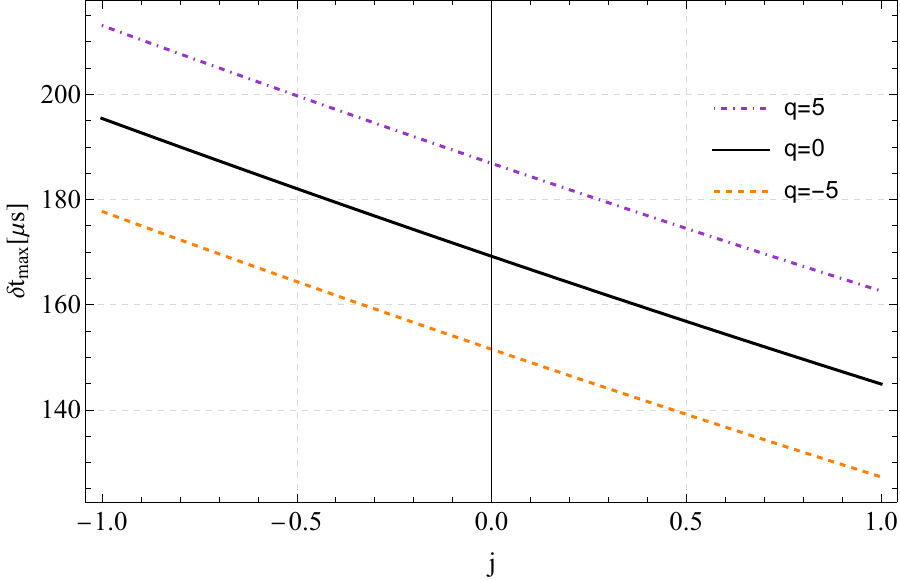}
\includegraphics[width=0.495\linewidth]{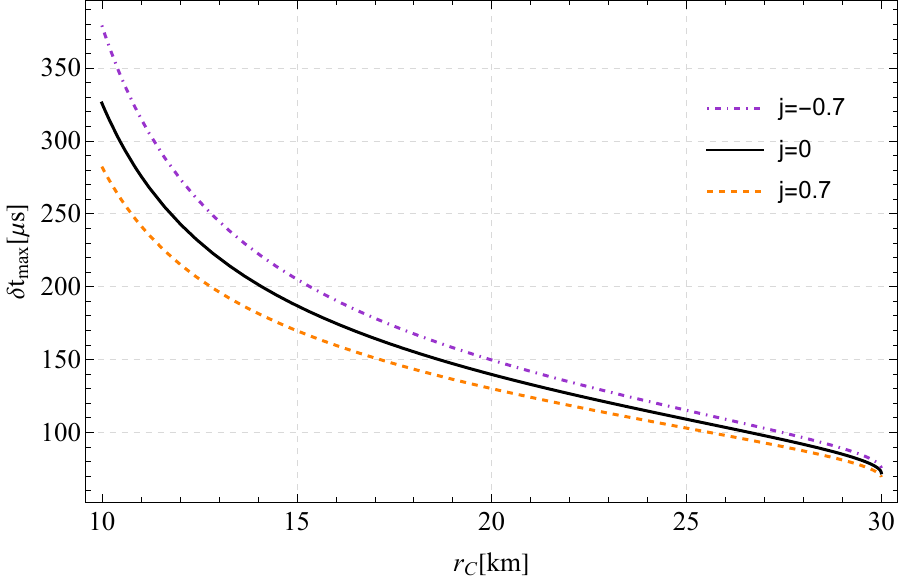}
\hfill} 
\caption{The time delay for Hartle-Thorne with respect to the angular momentum  $j=J/M^2$ (left panel) when a radar signal travels
back and forth through the path close to the neutron star, for different values of $q$ and  $1.4M_\odot$ mass of neutron star. Time delay as a function of distance for fixed $q=5$ from source of gravity to propagating light path $r_{C}$ for $j=0$, prograde $j=0.7$ and retrograde $j=-0.7$ rotation of neutron star (right panel).}
\label{fig:td_round_vs_j_HT}
\end{figure*}
\begin{figure*}[htbp]
{\hfill
\includegraphics[width=0.495\linewidth]{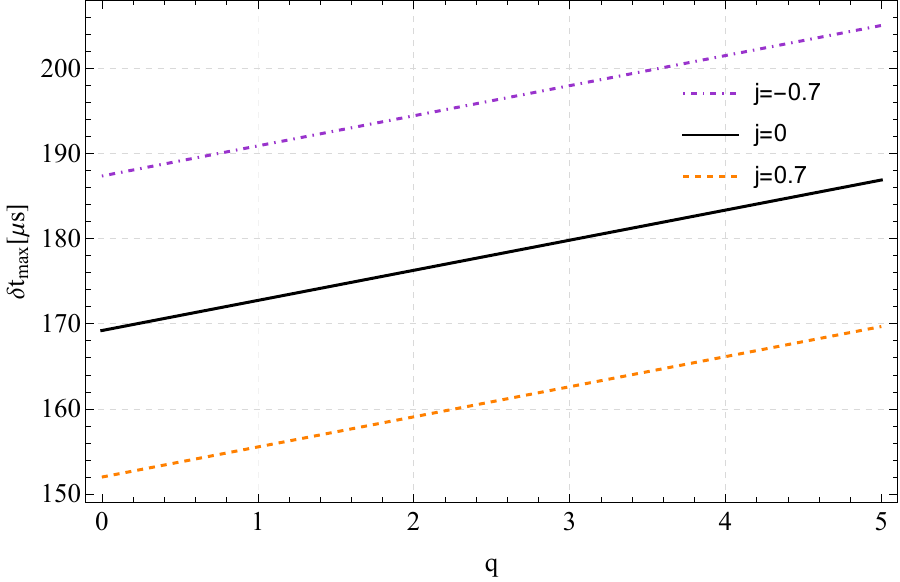}
\includegraphics[width=0.495\linewidth]{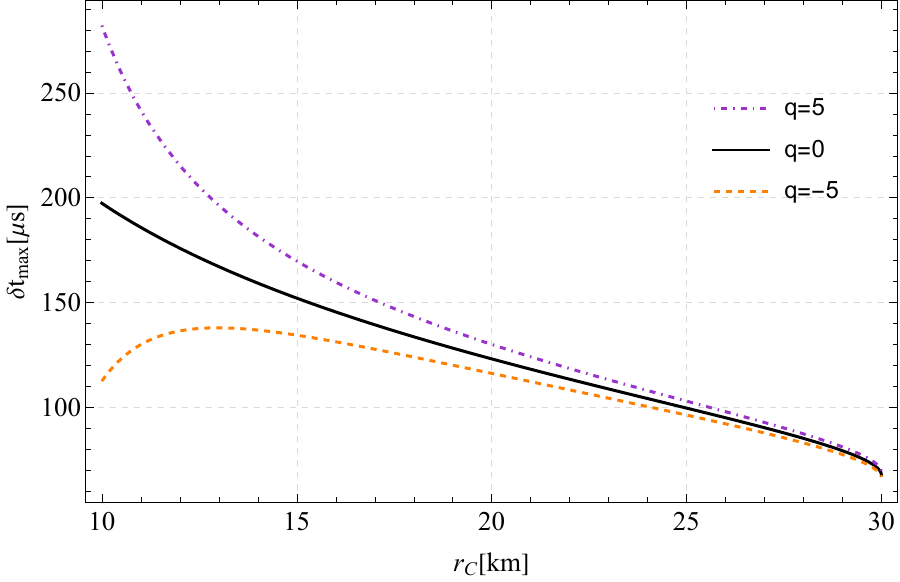}
\hfill} 
\caption{The time delay for Hartle-Thorne with respect to the deformation parameter $q=Q/M^3$ (left panel) when a radar signal travels
back and forth through the path close to the neutron star, when the exterior field is described by the Hartle-Thorne metric for different values of angular momentum and  $1.4M_\odot$ mass of neutron star. Time delay as a function of distance for fixed $j=0.7$ from source of gravity to propagating light path $r_{C}$, when $q=0$, oblate source $q=5$ and prolate $q=-5$ (right panel).}
\label{fig:td_round_vs_j_HT_q}
\end{figure*}

To fully analyze the Shapiro time delay in the Hartle-Thorne spacetime, we must consider the combined contributions of the spin terms $j$, $j^2$, and the quadrupole moment $q$. In this regime the analytical expressions become extremely lengthy, particularly with the quadratic terms $j^{2}$, so we present numerical results obtained by varying $j$ and $q$, using the full, non-expanded expressions throughout.

Figures~\ref{fig:numinttime_HT}--\ref{fig:td_round_vs_j_HT_q} summarize the behaviour of the one-way delay $t_{AC}$ (Fig.~\ref{fig:numinttime_HT}) and the round-trip excess $\delta t_{\max}$ (Figs.~\ref{fig:td_round_vs_j_HT} and \ref{fig:td_round_vs_j_HT_q}). Both grow with the mass $M$ and the quadrupole $q$, and decrease with the spin $j$ and with the closest-approach distance $r_C$. Rotation enters asymmetrically -- prograde ($j>0$) shortens the travel time, retrograde ($j<0$) lengthens it -- while an oblate source ($q>0$) increases the delay and a prolate source ($q<0$) reduces it; the slopes steepen with $M$ as stronger fields amplify both effects.

\begin{figure}
    \centering
    \includegraphics[width=1\linewidth]{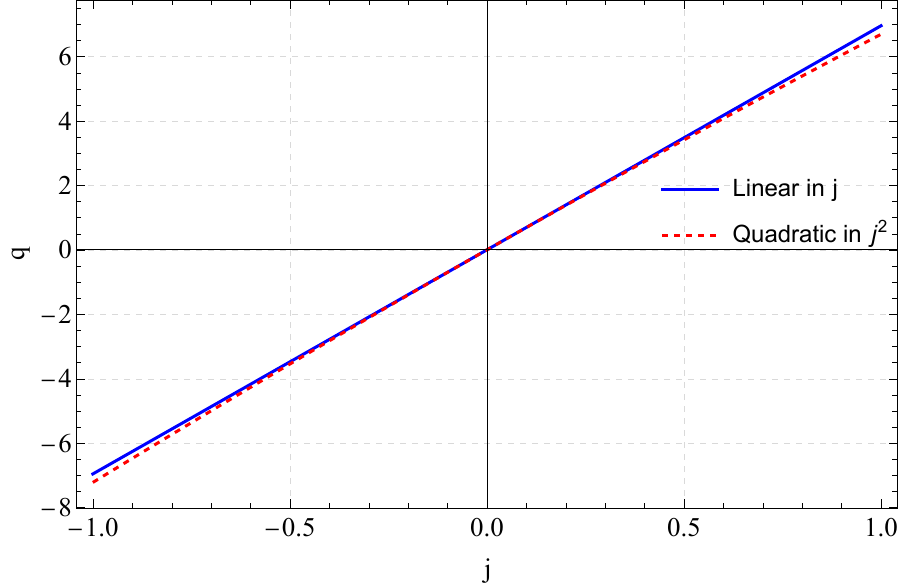}
    \caption{ Mimicking effect for the Shapiro delay in Hartle-Thorne metric for deformation parameter $q=Q/M^3$ and angular momentum $j=J/M^2$. The red dashed line corresponds to the Hartle-Thorne spacetime but the blue line is linear in $j$. 
    It illustrates how the quadrupole moment can effectively “mimic” contributions from angular terms in timing observables. The curves are plotted for a fixed orbital radius $r = 30$ km and stellar radius $r_C = 15$ km, with mass $1.4\,M_\odot$.}
    \label{fig:mim_Shap_HT}
\end{figure}

\begin{figure}
    \centering
    \includegraphics[width=1\linewidth]{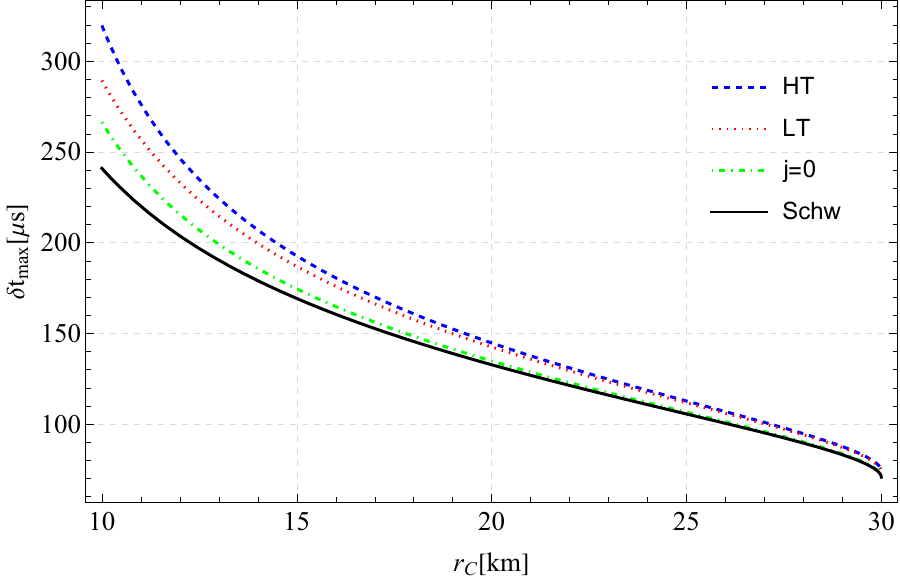}
    \caption{
Maximum time delay $\delta t_{\max}$ as a function of the orbital radius $r_C$ for four spacetime models: the Hartle-Thorne (HT), Lense-Thirring (LT), static deformed and Schwarzschild (Schw) solutions. The parameters are fixed to $q = 1.5$, $j =- 0.7$ and $M=1.4 M_\odot $. The HT metric (blue dashed curve) consistently predicts the largest time delay, followed by the LT approximation (red dotted curve), the HT metric in the $j \to 0$ limit with a deformation parameter (green dot-dashed curve), and the Schwarzschild case (black solid curve). All curves exhibit a monotonic decrease of $\delta t_{\max}$ with increasing radius, reflecting the weakening of relativistic effect at larger distances.}
    \label{fig:4plots}
\end{figure}

\subsection{Spin-quadrupole mimicking}\label{sec:shap_mimic}

Figure~\ref{fig:mim_Shap_HT} shows the mimicking effect within the Hartle-Thorne metric, including terms of order $j$, $j^2$, and $q$. In the slow-rotation regime ($|j|\ll1$), the inferred relation between $q=Q/M^3$ and $j=J/M^2$ is nearly linear: the linear frame-dragging term and the quadrupolar contribution enter the delay in the same combination, so a wide range of $(j,q)$ combinations produces almost identical time-delay profiles. For larger $|j|$, however, the quadratic term $j^2$ becomes important -- the linear and full curves diverge -- introducing curvature in the $q(j)$ relation that cannot be absorbed by a simple linear compensation. The degeneracy is therefore not limited to the Lense-Thirring sector but involves the full rotational structure of the Hartle-Thorne expansion.

Figure~\ref{fig:4plots} compares $\delta t_{\max}$ across four geometries at fixed $M$, $j$, and $q$. All curves decrease monotonically with $r_C$ and reveal a clear hierarchy: the Hartle-Thorne metric gives the largest delay (rotation and deformation combined), the Schwarzschild case the smallest (baseline), and the single-effect approximations -- Lense-Thirring (rotation only) and the static deformed limit -- fall in between. The convergence of all curves at large $r_C$ confirms that they approach the same weak-field limit.

\subsection{Weak-field limit: Solar-System estimate}\label{sec:shap_weakfield}
It is interesting to analyze the contribution of the spin parameter $j$ and quadrupole parameter $q$ to the gravitation field of the Sun. For that purpose, we use the total angular momentum $J$, the mass quadrupole $Q$, the gravitational constant $G$, and the speed of light in vacuum $c$. Then we expand the expression for the maximum time delay $\delta t_{max}$ with respect to the flat spacetime  when $r_C=R_\odot$ in Taylor series up to the terms $\sim 1/c^4$ in order to take into account the contribution of $J$. Eventually, bringing into mind that $R_\odot\ll r_A, r_B$, we obtain the following expression
\begin{equation} \label{eq:dtmax_sun}
\delta t_{max}=\frac{4GM}{c^3}\left(1+\ln\frac{4r_Ar_B}{r_C^2}\right)+\frac{12GQ}{c^3r_C^2}-\frac{16GJ}{c^4r_C},
\end{equation}
where the first term corresponds to the time delay due to the mass of the source \cite{2018igrc.book.....B}, the second term describes the shape of the source, and the last term accounts for the rotation of the source that induces the frame dragging effect. Now let us estimate the time delay of a radio signal going from Earth to Mercury and returning to Earth when the two planets are on opposite sides of the Sun. Using the numerical values of  $r_A=150 \times 10^6$ km, $r_B=58 \times 10^6$ km, $r_C=R_\odot=0.7 \times 10^6$ km, $M_\odot=2 \times 10^{30}$ kg, $Q_\odot\approx2.17\times10^{41}$ kg m$^2$ \cite{2000A&A...355..365G}, $J_\odot\approx1.92\times10^{41}$ kg m$^2$/s \cite{2012SoPh..281..815I} we got the following result:
\begin{eqnarray}
    \delta t_{max}(M)&=&0.24\times10^{-3} \, {\rm s} \, ,\\
    \delta t_{max}(Q)&=&1.33 \times10^{-11} \, {\rm s} \, ,\\
    \delta t_{max}(J)&=&-3.76 \times10^{-11} \, {\rm s} \, .
\end{eqnarray}
As one can observe, the main contribution to the delay time comes from the mass of the Sun, as expected. The contribution of $Q$ is smaller than that of $J$ in absolute value. Although the terms involving $Q$ and $J$ appear at different orders, $\sim 1/c^3$ and $\sim 1/c^4$, respectively, the frame-dragging effect is almost three times stronger than the effect of the Sun's quadrupolar deformation.
\subsection{Relation to pulsar timing and comparison with existing results}
\label{HTpulsar}

In the language of pulsar timing, the round-trip excess $\delta t_{\max}$
[Eqs.~\eqref{tACvsj}, \eqref{tACvsq} and, in the weak field,
\eqref{eq:dtmax_sun}] is the Shapiro contribution to the timing residual for a
signal propagating along the $A\!\to\!C\!\to\!B$ path of Fig.~\ref{fig:scheme}.

An analytic form of Shapiro delay to quadrupole order was obtained by Christian,
Psaltis \& Loeb~\cite{2016ApJ...820..139C} in the Butterworth--Ipser metric, as a
proof-of-principle no-hair test for a pulsar orbiting a spinning black hole,
using the time-transfer-function (iterative-integral) method for a general
orbital geometry. Our calculation is complementary: it is carried out in the
Hartle-Thorne neutron star exterior, where the quadrupole moment $Q$ is an
independent parameter, by direct integration of the null geodesics for the
equatorial two-body configuration of Fig.~\ref{fig:scheme}. Both results reduce
to the classical logarithmic mass delay and share the same expansion order, but
neither is a special case of the other.

A closer point of comparison is the recent work of Mora-Chaverri,
Santiago-Leandro \& Frutos-Alfaro~\cite{2025GReGr..57...55M}, who computed the
general relativistic time delay in the same Hartle-Thorne spacetime -- there
endowed additionally with a magnetic dipole and an electric charge -- from a
post-Newtonian form of the metric and by a different derivation of the delay. In
the regime where the two metric forms coincide (the $1/r$ expansion), and in the
common non-magnetic sector, our closed form expression agrees \emph{analytically}
with theirs, the magnetic dipole contribution they retain is about five orders of
magnitude smaller than the second order rotational term and does not affect the
comparison. This agreement, reached by an independent method, provides a direct
analytic check on our result.

We emphasise that this is a controlled analytic computation of the propagation
(Shapiro) term alone. A complete pulsar timing model would additionally require
the R\o{}mer and Einstein delays and the full orbital geometry of source and
observer, which lie beyond the present scope.

\section{Conclusions} \label{conclusions}

In this work we studied how rotation and quadrupole deformation jointly affect two relativistic observables -- the Shirokov and Shapiro effects -- in the Hartle-Thorne spacetime, the exterior field of a slowly rotating, axially symmetric neutron star.

Using the geodesic deviation formalism, we derived the equations governing the relative motion of neighbouring test-particle trajectories and found that the Shirokov effect depends clearly on the sign and magnitude of both $J$ and $Q$: the radial and azimuthal oscillations strengthen when both are present, enhancing the precession and modulation of the motion relative to the pure Lense-Thirring or purely static-quadrupolar cases. We also derived the Shapiro time delay in analytic form and analysed it in two limits. In the Lense-Thirring case ($Q=0$, $J^2=0$, $J\neq0$) frame dragging persists for both signs of the angular momentum, consistent with our earlier Lense-Thirring results; in the static-quadrupolar case ($J=0$, $J^2=0$, $Q\neq0$) more oblate sources produce a stronger delay with increasing distance, highlighting the role of deformation in shaping light propagation near neutron stars.

Both observables reveal a strong degeneracy between $j$ and $q$. In the low-spin regime different $(q,j)$ combinations yield nearly indistinguishable signatures, so that a slowly rotating neutron star can mimic a compact object of different multipole structure. Each observable therefore constrains only a combination of angular momentum and quadrupole and cannot separate them; breaking this degeneracy requires several independent measurements. Beyond the weak-field approximation, our full numerical analysis confirms this mimicking, and from its magnitude we find that the Shirokov effect contributes more strongly than the Shapiro effect.

A particularly noteworthy outcome concerns the weak-field limit of the Shirokov
effect. Although Shirokov originally regarded the splitting between the vertical
and in-plane oscillation periods as a purely relativistic phenomenon, our
evaluation for a realistic oblate source -- an Earth-orbiting satellite --
shows that the Newtonian quadrupole contribution exceeds the relativistic mass
term by roughly six orders of magnitude, and the relativistic quadrupole and
frame-dragging terms by two further orders. The relativistic Shirokov effect
therefore amounts to a small correction on top of a dominant Newtonian,
multipolar splitting. We develop this reinterpretation, and its derivation
directly from the Newtonian mass multipoles, in a companion
paper~\cite{idrissov2026newtonian}.

Comparison with previous analyses for the Lense-Thirring and Zipoy-Voorhees metrics shows consistent qualitative trends, with rotation and deformation modifying the frame-dragging and time delay signatures in a coherent way. This indicates that even moderate deviations from spherical symmetry can have measurable consequences for precision astrophysical observations, and that rotation and deformation must be treated jointly when interpreting timing signatures and relativistic precession around compact objects.

Natural extensions of this work include higher-order spin and quadrupolar contributions, observational consequences for pulsar timing and gravitational lensing in realistic neutron-star configurations, relativistic tidal interactions, and the gravitational-capture cross section for small bodies orbiting neutron stars. A similar analysis can be performed for black holes in alternative modified and extended theories of gravity \cite{2026JHEAp..4900425S,2025arXiv251103886S, 2025arXiv251016639S, 2025arXiv250916782S, 2025PDU....4801885E, 2025PDU....4801865A, 2024PDU....4601622E, 2025PDU....4801862B, 2025Univ...11...79Q, Idrissov:2017hdg, 2026arXiv260117081M}. These directions are left for future work.

\section*{Acknowledgments }

AI acknowledges financial support from SECIHTI through the National Postgraduate Scholarship Program (CVU 2222058). The work of HQ was supported by UNAM-DGAPA-PAPIIT, grant No. 108225, and Conahcyt, grant No. CBF-2025-I-243. 

\section*{Appendix}
\appendix
%
%
\allowdisplaybreaks

\section{Geodesic deviation coefficients}\label{app:gde_coeffs}
Here we collect the coefficients of the coupled deviation equations
\eqref{eq2}--\eqref{eq5}, expanded through $\mathcal{O}(j^{2})$ and
$\mathcal{O}(q)$ as in Eq.~\eqref{eq:abc-expansion}.
\color{black}
The coefficient $a_1$ takes the form:
\begin{equation}\label{eq6}
a_{1} = a_{10}(1+ja_{11}+j^{2}a_{12}+qa_{13}),
\end{equation}
where
\begin{align*}
a_{10}&=-2r u^{3}f,\\
a_{11}&=\frac{u^{0}M^{2}}{r^3 u^{3}},\\
a_{12}&=\frac{1}{16 M^2 f r^5}\Bigg[15 r^4 \left(4 M^3-6 M^2 r+r^3\right) \ln f\\
&\quad -2 M \big(48 M^6-64 M^5 r+12 M^4 r^2\\
&\quad +70 M^2 r^4-15 M r^5-15 r^6\big)\Bigg],\\
a_{13}&=\frac{5}{16 M^2 rf}\Bigg[28 M^3-6 M^2 r\\
&\quad -3 \left(4 M^3-6 M^2 r+r^3\right) \ln f-6 M r^2\Bigg].
\end{align*}
The coefficient $a_2$ is given by
\begin{equation}\label{eq7}
a_{2} = a_{20}(1+ja_{21}+j^{2}a_{22}+qa_{23}),
\end{equation}
where
\begin{align*}
a_{20}&=\frac{2M u^{0}f}{r^2},\\
a_{21}&=-\frac{u^{3}M}{u^{0}},\\
a_{22}&=-\frac{1}{16 M^3 r^3f}\Bigg[2 M \big(16 M^5+18 M^4 r\\
&\quad +30 M^3 r^2-25 M^2 r^3-30 M r^4\\
&\quad +15 r^5\big)+15 r^3 (r-2 M)^2 (M+r) \ln f\Bigg],\\
a_{23}&=\frac{5}{16 M^3f r^{2}}\Bigg[2 M (6 M^4 + 6 M^3 r\\
&\quad - 5 M^2 r^2 -6 M r^3 + 3 r^4)\\
&\quad + 3 r^2 (r-2 M)^2 (M + r) \ln f\Bigg].
\end{align*}
The coefficient $a_3$ can be expressed as
\begin{equation}\label{eq8}
a_{3} = a_{30}(1+ja_{31}+j^{2}a_{32}+q a_{33}),
\end{equation}
with
\begin{align*}
a_{30}&=\frac{2 M (u^{0})^2 (3 M-r)}{r^4}-(u^{3})^2,\\
a_{31}&=\frac{4 u^{0} u^{3}  (r-3 M)M^{2}}{2 M (u^{0})^2 (3 M-r)-r^4 (u^{3})^2},\\
a_{32}&=\frac{1}{16 M^2 r^2 \big(2 M (3 M - r) (u^{0})^2 - r^4 (u^{3})^2\big)}\\
&\quad \times\Bigg[2 M (80 M^5 +72 M^4 r + 90 M^3 r^2 - 20 M^2 r^3\\
&\qquad - 15 M r^4 - 15 r^5) (u^{0})^2 -6 M r (64 M^6\\
&\qquad -64 M^5 r + 8 M^4 r^2 - 10 M^2 r^4 + 15 M r^5\\
&\qquad +15 r^6) (u^{3})^2 +15 \big(-r^3 (-8 M^3 + r^3) (u^{0})^2\\
&\qquad -3 r^6 (r^2-2 M^2) (u^{3})^2\big) \ln f\Bigg],\\
a_{33}&=-\frac{5}{16 M^2 \left(2 M r (u^{0})^2 (r-3 M)+r^5 (u^{3})^2\right)}\\
&\quad \times\Bigg[6 M r^4 (u^{3})^2\left(-2 M^2+3 M r+3 r^2\right)\\
&\qquad +3 r^2 \ln f \big(-8 M^3 (u^{0})^2-6 M^2 r^3 (u^{3})^2\\
&\qquad +3 r^5 (u^{3})^2+r^3 (u^{0})^2\big)+2 M (u^{0})^2\\
&\qquad \left(-24 M^4-18 M^3 r+4 M^2 r^2+3 M r^3+3 r^4\right)\Bigg].
\end{align*}
The coefficient $b$ is
\begin{equation}\label{eq9}
b = b_{0}(1+j b_{1}+j^{2} b_{2}+q b_{3}),
\end{equation}
with
\begin{align*}
b_{0}&=\frac{2 u^{3}}{r},\\
b_{1}&=\frac{u^{0}M^{2}}{u^{3}r^{3} f},\\
b_{2}&=\frac{1}{16 M^2 r^4} \Bigg[ \frac{1}{r-2 M}\,2 M \big(64 M^6-56 M^5 r\\
&\quad -12 M^4 r^2-10 M^3 r^3-10 M^2 r^4-15 M r^5\\
&\quad +15 r^6\big)+15 r^6 \ln f\Bigg],\\
b_{3}&= \frac{5}{16 M^{2}fr^{2}}\Bigg[2 M (2 M^3 + 2 M^2 r\\
&\quad +3 M r^2 - 3 r^3) - 3 (r-2 M) r^3 \ln f\Bigg].
\end{align*}
The coefficient $c$ is
\begin{equation}\label{eq10}
c =c_{0}(1+j c_{1}+j^{2} c_{2}+qc_{3}),
\end{equation}
where
\begin{align*}
c_{0}&=\frac{2M u^{0}}{r^{2}f},\\
c_{1}&=-\frac{3 u^{3}M}{ u^{0}},\\
c_{2}&=\frac{1}{16 M^3 f r^5}\Bigg[2 M (64 M^7 - 64 M^6 r\\
&\quad - 16 M^5 r^2+ 2 M^4 r^3 +10 M^3 r^4 - 65 M^2 r^5\\
&\quad +60 M r^6 - 15 r^7) + 15 (M - r) r^5 (r-2 M )^2 \ln f \Bigg],\\
c_{3}&=\frac{5}{16 M^3 f r^{2}} \Bigg[2 M (2 M^4 - 2 M^3 r\\
&\quad +13 M^2 r^2 - 12 M r^3 +3 r^4)\\
&\quad -3 (M - r) r^2 (r-2 M)^2 \ln f\Bigg].
\end{align*}
Finally, the coefficient $h$ is given by
\begin{equation}\label{hcoef}
h=h_{0}(1+j h_{1}+j^{2} h_{2}+q h_{3}),
\end{equation}
where
\begin{align*}
h_{0}&=(u^{3})^{2},\\
h_{1}&=-\frac{4 u^{0}M^{2}}{u^{3}r^{3}},\\
h_{2}&=\frac{1}{16 M^2 r^7 (u^{3})^2} \Bigg[2 M (-48 M^6 + 8 M^5 r + 24 M^4 r^2\\
&\quad +30 M^3 r^3 + 60 M^2 r^4 - 135 M r^5 + 45 r^6) (u^{0})^2\\
&\quad +6 M r^3 (16 M^5 + 8 M^4 r - 10 M^2 r^3 + 15 M r^4 + 15 r^5) (u^{3})^2\\
&\quad +45 r^5 \big((-2 M + r)^2 (u^0)^2 + r^2 (-2 M^2 + r^2) (u^{3})^2\big)\ln{f}  \Bigg],\\
h_{3}&=\frac{15}{16 M^2 r^4 (u^3)^2}\Bigg[-2 M (M - r) (2 M^2 + 6 M r\\
&\quad -3 r^2) (u^{0})^2+2 M r^3 (2 M^2 - 3 M r - 3 r^2) (u^{3})^2\\
&\quad -3 r^2 \big((-2 M + r)^2 (u^{0})^2 + r^2 (-2 M^2 + r^2) (u^{3})^2\big) \ln{f} \Bigg].
\end{align*}
\section{Circular-geodesic coefficients}\label{app:circ_coeffs}

The functions entering the circular orbit condition
\eqref{explgeodeq_compact}, the normalisation \eqref{normaleq}, and the
four velocity components \eqref{equ0}--\eqref{equ3} are listed below.
\begin{align*}
K(r) &= \frac{2 M^{2}(2M - r)}{r^{3}}, \\[4pt]
C_{0}(r) &= \frac{1}{16 M^{2} r^{5}}\Big[
-15 r^{3}(r-2M)^{2}(M+r)\ln f\\
&\quad -2 M\,(16 M^{5} + 18 M^{4} r + 30 M^{3} r^{2}\\
&\quad -25 M^{2} r^{3} -30 M r^{4} +15 r^{5})\Big],\\[4pt]
C_{3}(r) &= \frac{1}{16 M^{2} r^{5}}\Big[
-15 r^{5}\big(4M^{3}-6M^{2}r + r^{3}\big)\ln f\\
&\quad +2M r\,(48 M^{6} -64 M^{5} r\\
&\quad +12 M^{4} r^{2} +70 M^{2} r^{4} -15 M r^{5} -15 r^{6})\Big],\\[4pt]
D_{0}(r) &= \frac{5}{16 M^{2} r^{4}}\Big[
3 r^{2} (r-2M)^{2}(M+r) \ln f\\
&\quad +2 M\,(6 M^{4} +6 M^{3} r -5 M^{2} r^{2} -6 M r^{3} +3 r^{4})\Big],\\[4pt]
D_{3}(r) &= \frac{5}{16 M^{2} r^{4}}\Big[
3 r^{4} \big(4M^{3}-6M^{2}r + r^{3}\big) \ln f\\
&\quad +2 M r^{4}\,(-14 M^{2} + 3 M r + 3 r^{2})\Big],\\[4pt]
A_{0}(r) &= \frac{1}{16 M^{2} r^{5}}\Big[15 r^{5}(r-2M)^{2}\ln f\\
&\quad -2M(M-r)\big(16M^{5}+8M^{4}r\\
&\quad -10M^{2}r^{3}-30Mr^{4}+15r^{5}\big)\Big],\\[4pt]
A_{3}(r) &= \frac{1}{16 M^{2} r^{5}}\Big[15 r^{7}(r^{2}-2M^{2})\ln f\\
&\quad +2Mr^{3}\big(16M^{5}+8M^{4}r\\
&\quad -10M^{2}r^{3}+15Mr^{4}+15r^{5}\big)\Big],\\[4pt]
B_{0}(r) &= \frac{5}{16 M^{2} r^{2}}\Big[-3 r^{2} (r - 2M)^{2} \ln f\\
&\quad - 2M (M - r)\left(2 M^{2} + 6 M r - 3 r^{2}\right)\Big],\\[4pt]
B_{3}(r) &= \frac{5}{16 M^{2} r^{2}}\Big[-3 r^{4} (r^{2} - 2M^{2}) \ln f\\
&\quad + 2M r^{3}\left(2 M^{2} - 3 M r - 3 r^{2}\right)\Big].
\end{align*}
\begin{align*}
u^{01}&=\left(1-\frac{3 M}{r}\right)^{-1/2},\\
g^{00}&=-\frac{3 M^{5/2} }{r^{3/2} (r-3 M)},\\
g^{1}&=\frac{1}{32\,M\,r^{6}\,(2M-r)\,\bigl(1-\tfrac{3M}{r}\bigr)^{2}}\\
&\quad \times\Bigg[2 M \big(144 M^{7}-192 M^{6} r+108 M^{5} r^{2}-74 M^{4} r^{3}\\
&\qquad -80 M^{3} r^{4}+345 M^{2} r^{5}-240 M r^{6}+45 r^{7}\big)\\
&\qquad -15 r^{5}\big(24 M^{3}-38 M^{2} r+19 M r^{2}-3 r^{3}\big)\,\ln f\Bigg],\\
g^{2}&=\frac{15}{32\,M\,r\,(6M^{2}-5Mr+r^{2})}\\
&\quad \times \Bigg[2M\bigl(2M^{3}+2M^{2}r-7Mr^{2}+3r^{3}\bigr)\\
&\qquad +r^{2}\bigl(8M^{2}-10Mr+3r^{2}\bigr)\ln f\Bigg],\\
u^{03}&=\frac{M^{1/2}}{r (r-3M)^{1/2}},\\
d^{03}&=-\frac{ M^{3/2} }{r^{1/2}  (r-3 M)},\\
d^{3}&=\frac{1}{32 M^{3} r^4 (r-3 M)^{2}}\Bigg[-15 r^4 (r-3 M)\\
&\quad \times\left(6 M^4-6 M^3 r+3 M^2 r^2-3 M r^3+r^4\right) \ln{f}\\
&\quad -2 M (144 M^8-192 M^7 r+12 M^6 r^2-66 M^5 r^3\\
&\qquad +243 M^4 r^4-135 M^3 r^5+110 M^2 r^6\\
&\qquad -75 M r^7+15 r^8)\Bigg],\\
d^{4}&=\frac{5}{32 M^{3} r (r-3 M)}\\
&\times \Bigg[2 M \left(6 M^4-15 M^3 r+4 M^2 r^2-6 M r^3+3 r^4\right)\\
&\quad +3 r \left(6 M^4-6 M^3 r+3 M^2 r^2-3 M r^3+r^4\right) \ln{f}\Bigg].
\end{align*}
\section{Frequency and period coefficients}\label{app:freq_coeffs}

The second-order spin and quadrupole coefficients of the vertical and in-plane
frequencies, Eqs.~\eqref{bigfreq} and \eqref{smallfreq}, and of the period
ratio \eqref{TroverTtheta}, are given below.
\begin{align*}
\Omega_{03}&=-\frac{1}{16 M^3 r^4 (3 M-r)^2}  \Bigg[2 M \big(144 M^8\\
&\quad -192 M^7 r-12 M^6 r^2+6 M^5 r^3-726 M^4 r^4+1305 M^3 r^5\\
&\quad -940 M^2 r^6+285 M r^7-30 r^8\big) - 15 r^4 \big(r-3 M\big)\\
&\quad \times \big(12 M^4-36 M^3 r+36 M^2 r^2-15 M r^3+2 r^4\big) \ln{f}\Bigg],\\[6pt]
\Omega_{04}&=\frac{5}{16 M^3 r (r-3 M)}\Bigg[2 M \big(6 M^4\\
&\quad +48 M^3 r-71 M^2 r^2+39 M r^3-6 r^4\big)-3 r \big(12 M^4\\
&\quad -36 M^3 r+36 M^2 r^2-15 M r^3+2 r^4\big) \ln{f}\Bigg].
\end{align*}
\begin{align*}
\omega_{03}&=-\frac{1}{16 M^3 r^5 f (r-3 M)^2 (r-6M)}\\
&\quad \times \Bigg[2 M \big(2592 M^{10}-4896 M^9 r+2280 M^8 r^2\\
&\qquad -1116 M^7 r^3-204 M^6 r^4-2568 M^5 r^5+8142 M^4 r^6\\
&\qquad -7825 M^3 r^7+3470 M^2 r^8-735 M r^9+60 r^{10}\big)\\
&\qquad +15 r^6 (r-3 M) \big(12 M^4\\
&\qquad -90 M^3 r+90 M^2 r^2-33 M r^3+4 r^4\big) f\ln{f}\Bigg],\\[6pt]
\omega_{04}&= \frac{5}{16 M^3 r^2 f (r-3 M) (r-6M)}\\
&\quad \times \Bigg[-3 r^2 \big(24 M^5-192 M^4 r+270 M^3 r^2\\
&\qquad - 156 M^2r^3 +41 M r^4 - 4 r^5\big) \ln{f}\\
&\qquad +2 M \big(108 M^6+12 M^5 r+180 M^4 r^2\\
&\qquad -482 M^3 r^3+361 M^2 r^4-111 M r^5+12 r^6\big)\Bigg].
\end{align*}
\begin{align*}
\left(\frac{T_r}{T_\theta}\right)_{03}
&= \frac{3}{16 M^3 (2M - r) r^4 (r - 6M)^2}\\
&\quad \times \Bigg[2M \big(96 M^9 - 496 M^8 r + 344 M^7 r^2\\
&\qquad - 96 M^6 r^3+ 854 M^5 r^4 - 2218 M^4 r^5\\
&\qquad + 2265 M^3 r^6 - 1040 M^2 r^7 + 210 M r^8 - 15 r^9\big)\\
&\qquad - 15 r^4 (r - 6M) (r - 2M)^2\\
&\qquad \times \left(-2 M^3 + 5 M^2 r - 5 M r^2 + r^3\right)\ln{f}\Bigg],\\[6pt]
\left(\frac{T_r}{T_\theta}\right)_{04}
&= \frac{1}{16 M^3 r (12 M^2 - 8 M r + r^2)}\Bigg[30 M (M - r)\\
&\quad \times \left(2 M^4 - 26 M^3 r + 43 M^2 r^2- 21 M r^3 + 3 r^4 \right)\\
&\quad - 45 r (r - 2M)^2\\
&\quad \times\left(-2 M^3 + 5 M^2 r - 5 M r^2 + r^3\right)\ln{f}\Bigg].
\end{align*}
\section{Shapiro-delay integrand coefficients}\label{app:shapiro_coeffs}
The second-order spin and quadrupole coefficients of the integrand
\eqref{eq:integrandgen}, together with the auxiliary polynomial, are collected
\begin{align} \nonumber
\left(\frac{1}{\sqrt{I}}\right)_2
&= \frac{1}{16 M^2 r^4 (r - 2M)(r - r_C)\, r_C (r_C - 2M)\, \Delta^2}
\nonumber \\
&\times \Bigg[
2M (r - r_C)\, \mathcal{P}(r, r_C)
\nonumber \\
&\quad
- 15 r^4 (r - 2M)^2 (r_C - 2M)\, r_C \,\Delta
\nonumber \\
&\qquad \times
\left[2 M r^4 - r^4 r_C - (M^2 + 3Mr - 2r^2) r_C^3\right]
\nonumber \\
&\qquad \times \ln\!\left(1 - \frac{2M}{r}\right)
\nonumber \\
&\quad
- 15 r^4 (r - 2M)^2 r_C^4 \,\Delta
\nonumber \\
&\qquad \times
\left(2 M^3 + M^2 r_C - 3 M r_C^2 + r_C^3\right)
\nonumber \\
&\qquad \times \ln f_C
\Bigg], \nonumber
\end{align}
\begin{align} \nonumber
\left(\frac{1}{\sqrt{I}} \right)_3
&= \frac{5 }{32 M^2 r (r - 2M)(r_C - 2M)(r - r_C)\, \Delta}
\nonumber \\
&\times \Bigg[
-4 M (r - r_C)
\nonumber \\
&\quad \times \Big(
3 r^3 r_C^2 (r^2 + r r_C - r_C^2)
\nonumber \\
&\qquad
+ 4 M^5 (2 r^2 + 2 r r_C + 5 r_C^2)
\nonumber \\
&\qquad
- M^3 r (36 r^3 + 52 r^2 r_C + 47 r r_C^2 - 25 r_C^3)
\nonumber \\
&\qquad
+ 2 M^4 (8 r^3 + 4 r^2 r_C - 2 r r_C^2 - 5 r_C^3)
\nonumber \\
&\qquad
- 3 M r^2 r_C (4 r^3 + 7 r^2 r_C + 3 r r_C^2 - 4 r_C^3)
\nonumber \\
&\qquad
+ 4 M^2 r (3 r^4 + 12 r^3 r_C + 13 r^2 r_C^2
\nonumber \\
&\qquad\qquad
- r r_C^3 - 3 r_C^4)
\Big)
\nonumber \\
&\quad
- 6 r (r - 2M)^2 (r_C - 2M)
\nonumber \\
&\qquad \times
\left(-2 M r^4 + r^4 r_C + M^2 r_C^3
+ 3 M r r_C^3 - 2 r^2 r_C^3\right)
\nonumber \\
&\qquad \times \ln f
\nonumber \\
&\quad
+ 6 r (r - 2M)^2 r_C^3
\left(2 M^3 + M^2 r_C - 3 M r_C^2 + r_C^3\right)
\nonumber \\
&\qquad \times \ln f_C
\Bigg], \nonumber
\end{align}
where
\begin{equation} \nonumber
\Delta = r\,r_C\,(r+r_C) - 2M(r^2 + r r_C + r_C^2),
\end{equation}
\begin{equation}\nonumber
f_C =1 -\frac{2 M}{r_C},
\end{equation}
\begin{equation}\label{Ppoly} \nonumber
\mathcal{P}(r,r_C)=\sum_{k=0}^{7} p_k\,r_C^{\,k},
\end{equation}
with coefficients
\begin{align}\nonumber
p_0 ={}& 48 M^7 r^5 (r-2M)^2,\nonumber\\[2pt]
p_1 ={}& 8 M^3 r^4\big(16 M^6-48 M^5 r+52 M^4 r^2\nonumber\\
&\quad -22 M^3 r^3-20 M^2 r^4+45 M r^5\nonumber\\
&\quad -15 r^6\big),\nonumber\\[2pt]
p_2 ={}& 4 M^2 r^3\big(16 M^7-48 M^6 r+72 M^5 r^2\nonumber\\
&\quad -84 M^4 r^3-45 M^3 r^4+240 M^2 r^5\nonumber\\
&\quad -195 M r^6+45 r^7\big),\nonumber\\[2pt]
p_3 ={}& -2 M r^2 (r-2M)^2\big(24 M^6-12 M^5 r\nonumber\\
&\quad +4 M^4 r^2+41 M^3 r^3+60 M^2 r^4\nonumber\\
&\quad -135 M r^5+45 r^6\big),\nonumber\\[2pt]
p_4 ={}& r (r-2M)\big(64 M^8-176 M^7 r+232 M^6 r^2\nonumber\\
&\quad +52 M^5 r^3-6 M^4 r^4-395 M^3 r^5\nonumber\\
&\quad +530 M^2 r^6-195 M r^7+15 r^8\big),\nonumber\\[2pt]
p_5 ={}& 2 (r-2M)^2\big({-8} M^7+20 M^6 r-28 M^5 r^2\nonumber\\
&\quad -9 M^4 r^3+5 M^3 r^4+30 M^2 r^5\nonumber\\
&\quad -30 M r^6+15 r^7\big),\nonumber\\[2pt]
p_6 ={}& M (r-2M)^2\big(8 M^5-12 M^4 r+16 M^3 r^2\nonumber\\
&\quad +25 M^2 r^3-50 M r^4+15 r^5\big),\nonumber\\[2pt]
p_7 ={}& -15 r^4 (r-2M)^3. \nonumber
\end{align}
\section{Shapiro delay in the weak-field limit}\label{app:weakfield_shapiro}
%
%
\allowdisplaybreaks
In this appendix we derive the Shapiro time delay in the weak-field limit of the
Hartle--Thorne metric, obtained by expanding the integrand
\eqref{eq:integrandgen} to first order in $M$. For compactness we introduce the
radial combinations
\begin{equation*}
\rho \equiv r+r_C,\qquad \sigma \equiv r^{2}+r r_C+r_C^{2},
\end{equation*}
and the endpoint combinations
\begin{equation*}
D \equiv \sqrt{r_A^{2}-r_C^{2}},\quad \Sigma \equiv r_A+r_C,\quad
\Theta \equiv \arctan\!\frac{r_C}{D}.
\end{equation*}
The expanded integrand then reads
\begin{equation}\label{I3}
\begin{split}
\frac{1}{\sqrt{I}}\approx{}&
\sqrt{\frac{r^{2}}{r^{2}-r_C^{2}}}\,
\bigg[\,1+\frac{M(2r+3r_C)}{r\,\rho}\\
&\;\pm J\,\mathcal{I}_{J}+J^{2}\,\mathcal{I}_{J^{2}}+Q\,\mathcal{I}_{Q}\,\bigg],
\end{split}
\end{equation}
with coefficients
\begin{align*}
\mathcal{I}_{J} &= \frac{2\,\sigma\,(3M\sigma+r r_C\rho)}{r^{4} r_C\,\rho^{2}},\\
\mathcal{I}_{J^{2}} &= \frac{1}{28 r^{5}}\bigg[
\frac{318 M r^{3}}{r_C^{3}}
+\frac{105 r^{3}}{r_C^{2}}
+\frac{1641 M r}{r_C}\\
&\quad -477 M-91 r
+\frac{2r}{\rho^{3}}\Big(84 r\rho(M+r)\\
&\quad -775 M\rho^{2}-420 M r^{2}\Big)\bigg],\\
\mathcal{I}_{Q} &= \frac{1}{4 r^{4}}\bigg[
47 M+8 r
+\frac{19 M r^{2}}{r_C^{2}}
+\frac{12 M r^{2}}{\rho^{2}}\\
&\quad -\frac{4 r(2M+r)}{\rho}
+\frac{4 r^{2}}{r_C}\bigg].
\end{align*}
After integration from $r_{C}$ to $r_A$ we obtain
\begin{equation}\label{tAC_HT}
\begin{split}
t_{AC}={}&2M\ln\!\frac{D+r_A}{r_C}\\
&+D\Big[\,1+\frac{M}{\Sigma}
\pm J\,\mathcal{T}_{J}
+J^{2}\,\mathcal{T}_{J^{2}}
+Q\,\mathcal{T}_{Q}\,\Big],
\end{split}
\end{equation}
where
\begin{align*}
\mathcal{T}_{J} &= \frac{1}{2\Sigma^{2}}\bigg[
12+\frac{4r_C}{r_A}+\frac{8r_A}{r_C}\\
&\quad +\frac{12M}{r_A}-\frac{14M}{r_C}
+\frac{6M r_C}{r_A^{2}}-\frac{16M r_A}{r_C^{2}}\\
&\quad +\frac{15\pi M\Sigma^{2}}{r_C^{2} D}
+\frac{30 M D\,\Sigma\,\Theta}{r_C^{2}(r_C-r_A)}\bigg],\\
\mathcal{T}_{J^{2}} &= \frac{1}{112}\bigg[
\frac{70(135M-13r_C)\,\Theta}{r_C^{4} D}\\
&\quad -\frac{1}{\Sigma^{3}}\Big(
\frac{1666}{r_C}+\frac{546}{r_A}
+\frac{2198 r_A}{r_C^{2}}\\
&\qquad +\frac{182 r_C}{r_A^{2}}
+\frac{896 r_A^{2}}{r_C^{3}}
+\frac{1726 M}{r_A^{2}}\\
&\qquad -\frac{31870 M}{r_C^{2}}
-\frac{4238 M}{r_A r_C}
+\frac{636 M r_C}{r_A^{3}}\\
&\qquad -\frac{44166 M r_A}{r_C^{3}}
-\frac{18296 M r_A^{2}}{r_C^{4}}\\
&\qquad +\frac{4725\pi M\Sigma^{3}}{r_C^{4} D}
-\frac{455\pi\Sigma^{3}}{r_C^{3} D}
\Big)\bigg],\\
\mathcal{T}_{Q} &= \frac{1}{16}\bigg[
\frac{1}{\Sigma^{2}}\Big(
\frac{80}{r_C}+\frac{32}{r_A}
+\frac{48 r_A}{r_C^{2}}\\
&\quad +\frac{94 M}{r_A^{2}}
-\frac{82 M}{r_C^{2}}
+\frac{156 M}{r_A r_C}\\
&\quad -\frac{128 M r_A}{r_C^{3}}
+\frac{125\pi M\Sigma^{2}}{r_C^{3} D}\Big)
-\frac{250 M\,\Theta}{r_C^{3} D}\bigg].
\end{align*}
\newpage
\bibliography{0refs}
\end{document}